\documentclass[fleqn,usenatbib]{mnras}
\usepackage{newtxtext,newtxmath}
\usepackage[T1]{fontenc}
\DeclareRobustCommand{\VAN}[3]{#2}
\let\VANthebibliography\thebibliography
\def\thebibliography{\DeclareRobustCommand{\VAN}[3]{##3}\VANthebibliography}

\usepackage{soul}
\usepackage[dvipsnames]{xcolor}

\usepackage{diagbox}
\usepackage{graphicx}	
\usepackage{amsmath}	

\graphicspath{{figures/}}

\title[Dust rings are protoplanet traps]{Dust rings trap
  protoplanets on eccentric orbits and get consumed by them}

\author[Velasco-Romero et al.]{%
David A. Velasco-Romero$^{1}$\thanks{E-mail: david.velasco@princeton.edu},
  Fr\'ed\'eric S. Masset$^{2,3}$,
  Alessandro Morbidelli$^{3}$,
  \newauthor
   Pablo Ben\'\i tez-Llambay$^{4}$,
  Leonardo Krapp$^{5,6}$ and  Elena Lega$^{3}$\\
$^{1}$Department of Astrophysical Sciences, Princeton University, Princeton, NJ 08544, USA\\
$^{2}$Instituto de Ciencias F\'isicas, Universidad Nacional
Aut\'onoma de M\'exico, Av. Universidad s/n, 62210 Cuernavaca, Mor., Mexico\\
$^{3}$Universit\'e Nice-Sophia Antipolis, CNRS, Observatoire de la
C\^ote d'Azur, Laboratoire Lagrange, CS 34229, 06304 Nice Cedex,
France\\
$^{4}$Facultad de Ingenier\'ia y Ciencias, Universidad Adolfo 
Ib\'añez, Av. Diagonal las Torres 2640, Peñalol\'en, Chile\\
$^{5}$Department of Astronomy and Steward Observatory,  University of Arizona, Tucson, Arizona 85721, USA\\
$^{6}$Departamento de Astronomía,  Universidad de Concepción, Av. Esteban Iturra s/n Barrio Universitario, Casilla 160-C, Chile
}

\date{Accepted XXX. Received YYY; in original form ZZZ}

\pubyear{2024}

\begin{document}
\label{firstpage}
\pagerange{\pageref{firstpage}--\pageref{lastpage}}
\maketitle

\begin{abstract}
  We study the orbital evolution and mass growth of protoplanets with
  masses $M \in [0.1-8]$~M$_\oplus$ in the vicinity of a dusty ring,
  using three-dimensional numerical simulations with a two-fluid model
  and nested-meshes. We find two stable, eccentric orbits that lock
  the planet in the ring vicinity, thereby inhibiting its migration
  and allowing it to accrete dust from the ring. One of these orbits
  has an eccentricity comparable to the aspect ratio of the gaseous
  disc and has its periastron within the ring, enabling intermittent
  accretion during each pass. The other orbit has a smaller
  eccentricity and an apoastron slightly inside the ring. A planet
  locked at the outer orbit efficiently accretes from the ring and can
  reach the critical mass for runaway gas accretion on timescales
  $\gtrsim 10^5$ yr (for a 10~M$_\oplus$ dust ring at 10~au) while a
  planet locked at the inner orbit has a slower growth and might not
  supersede the super-Earth stage over the disc lifetime. While in our
  runs a low-mass embryo forming within the ring eventually joins the
  outer orbit, it is likely that the path taken depends on the
  specific details of the ring. The trapping on the outer orbit arises
  from an intermittent, strong thermal force at each passage through
  the ring, where the accretion rate spikes. It is insensitive
    to uncertainties that plague models considering planets
    trapped on circular orbits in rings. It is highly robust and could
    allow a growing planet to follow an expanding ring over large distances.
\end{abstract}



\section{Introduction}
\label{sec:introduction}
Dust rings are ubiquitous in continuum millimetre images of
protoplanetary discs obtained with ALMA
\citep{2018ApJ...869L..42H}. Even discs that appear smooth with
standard image reconstruction techniques reveal annular substructures
when super-resolution techniques are used
\citep{2022MNRAS.509.2780J}. Their origin is not known. While some are
thought to be the result of planet-disc interactions
\citep{2018ApJ...869L..47Z,2020A&A...637A..50Z}, others may have a different origin. Dust
is known to accumulate at local pressure maxima
\citep{1972fpp..conf..211W,2005MNRAS.364L..81F,2007ApJ...664L..55K,2012A&A...538A.114P}. These may occur, for
instance, in discs subject to non-ideal MHD effects
\citep{2017A&A...600A..75B,2018ApJ...865..105K}, and are accompanied by the formation of
thin, narrow rings \citep{2018A&A...617A.117R,
  2020A&A...639A..95R}. Not all explanations for the existence of
narrow rings resort to local pressure
maxima. When the pressure profile has a small perturbation, insufficient
  to create a local maximum, the radial drift of the dust slows down
  where the pressure gradient is small, resulting in a traffic jam
  of the inwardly flowing dust that appears as a dusty ring
  \citep{2016MNRAS.459.2790R}. Traffic jams (hence rings) may also
  appear at ice lines \citep{2017A&A...608A..92D}. \citet{2021MNRAS.505.1162J} introduce the concept of clumpy
rings, that can exist in discs where the pressure has a smooth, monotonous
profile. Clumpy rings are the results of radially localised
  formation of pebble clumps actively forming planetesimals, that are
  fed by the settling of pebbles and their incorporation into clumpy
  structures with dust-to-gas ratio in the midplane larger than
  unity. They may survive in the absence of a pressure bump, and
  require to be fed by an inward flow of solids from the outer disc,
  while they leak solids toward the inner disc at a
  sizeable fraction of the inflow rate from the outer disc. When
  the conditions for their survival are met, they usually expand
  outwards and can survive over evolutionary timescales of the disc.

  Given that dusty rings are sweet spots for the formation of
planetesimals, it is legitimate to investigate how the growth of a
planetary embryo would unfold at a dusty ring. In recent years, there
has been significant work addressing the growth
and orbital evolution of planetary bodies at dusty rings. We shall
present and discuss those in section~\ref{sec:comp-prev-work}. Here,
we take into account the radiative feedback due to the diffusion
into the ambient gas of the energy released by accretion of pebbles as
the planet passes through the ring.
Indeed, recent work has highlighted the important role played by
thermal disturbances on the orbital evolution of low-mass planet in
the vicinity of dusty rings \citep{2023MNRAS.524.2705C,
  2024arXiv240205760P}. In general, they tend to excite the planet's
eccentricity to values comparable to the aspect ratio of the gaseous
disc, so that their growth and orbital evolution considerably differs
from that obtained assuming a circular orbit at a migration trap in
the ring.

The force arising from these disturbances, or thermal force, can
dominate the force exerted by the disc for low-mass planets, to the point
that the classical Lindblad and corotation torques are largely
subdominant, if relevant at all. This happens when the thermal
lengthscale $\lambda$, given by:
\begin{equation}
  \label{eq:1}
  \lambda = \sqrt{\frac{\chi}{\frac 32\Omega_\mathrm{K}\gamma}}
\end{equation}
is much smaller than the pressure lengthscale $H=c_s/(\sqrt{\gamma}\Omega_\mathrm{K})$, where
$\Omega_K$ is the Keplerian angular speed, $\chi$ the thermal diffusivity,
$\gamma$ the adiabatic index and $c_s$ the adiabatic sound
speed. Estimates of the thermal lengthscale in planet forming regions
of protoplanetary discs, at a few astronomical units from the central
object, show that it is indeed a minute fraction of
the pressure lengthscale \citep{2017MNRAS.472.4204M}.

The action of thermal forces on an embedded planet with a mass $M$
  significantly smaller than the thermal mass $h^3M_\star$ ($M_\star$ being the mass
of the central object and $h$ the aspect ratio of the gaseous disc)
depends on
the planet's luminosity $L$. When the luminosity is larger than the
critical luminosity $L_c$ given by \citep{2017MNRAS.472.4204M}
\begin{equation}
  \label{eq:2}
  L_c=\frac{4\pi GM\chi\rho_0}{\gamma},
\end{equation}
where $G$ is the gravity constant and $\rho_0$
the gas density at the disc's midplane, thermal forces tend to induce
an outward migration \citep{2015Natur.520...63B} for circular or low
eccentricity planets, while at the same time they tend to excite their
eccentricity
\citep{2017A&A...606A.114C,2017arXiv170401931E,2019MNRAS.485.5035F,2022MNRAS.509.5622V,2023MNRAS.tmp..658C}. As
the eccentricity reaches values smaller, but comparable to the aspect
ratio of the disc, the migration is found to revert inwards
\citep{2017arXiv170401931E,2023MNRAS.524.2705C}. An early
interpretation of this reversal was put forward by
\citet{2017arXiv170401931E}, who speculated that the (positive)
corotation torque was quenched by the relatively large value of the
eccentricity \citep{2014MNRAS.437...96F}, and that this effect
sufficed to revert the torque balance. Regardless of its origin, this
reversal can have a decisive effect on the growth and orbital
evolution of a planet: should it occur systematically for planets
becoming eccentric at dusty rings, they would migrate inwards and
leave the ring, thereby interrupting their growth at a very low mass
\citep{2023MNRAS.524.2705C}.

This paper is organised as follows: In
section~\ref{sec:torq-revers-diff}, we revisit this reversal and shed
some light on its physical origin.  We show that it is intrinsic to,
and explained entirely by the behaviour of the thermal force. This, in
turn, will lead us to conceive situations in which migration is not
reverted at sizeable eccentricity. We shall see that such conditions
may easily be met at the edges of dusty rings. We subsequently resort
to numerical simulations to check our expectations. The setup is
presented in section~\ref{sec:numer-simul} and the results in
section~\ref{sec:results}. We then discuss our results in
section~\ref{sec:discussion} and conclude in
section~\ref{sec:conclusions}.

\section{On the thermal torque reversal with eccentricity}
\label{sec:torq-revers-diff}
Let us start by noting that, \emph{when the eccentricity is constant
  in time}, the time averaged torque can be used to infer the
migration rate. The time derivative of the planet's angular momentum
$J=M\sqrt{GM_\star a(1-e^2)}$ ($a$ and $e$ being the semi-major axis and eccentricity of the planet) is indeed:
\begin{equation}
  \label{eq:3}
  \Gamma = \dot J = J\left(\frac{\dot
      a}{2a}-\frac{e\dot e}{1-e^2}\right),
\end{equation}
hence the net torque $\Gamma$ exerted on the planet scales directly
with the migration rate $\dot a$ when the second term in the
parenthesis of the equation above vanishes.

We therefore proceed to evaluate the thermal torque exerted on a
low-mass planet, averaged over one orbital period. We entertain two
cases: (i) a regime with a radial excursion smaller than the thermal
lengthscale, and (ii) a regime with a radial excursion larger than the
thermal lengthscale, but smaller than the aspect ratio. This case can
exist when the thermal and pressure length scales are well separated.

Case (i) has been studied in detail by \citet{2019MNRAS.485.5035F}. We
adopt notations similar to theirs. The planet has coordinates $(x,y)$
in a frame centred on the guiding centre of the epicycle ($x$ being
directed along the radial direction and $y$ by the azimuthal direction):
\begin{equation}
  \begin{aligned}
  \label{eq:4}
  x&=-ea\cos(\Omega_pt)\\
  y&=2ea\sin(\Omega_pt),
  \end{aligned}
\end{equation}
where $\Omega_p$ is the planet's orbital frequency and $t$ the time
since a passage at periastron. We work out the thermal force exerted
on the planet to first order in eccentricity. The constant term is the
force exerted on a planet in circular orbit, and has expression
\citep[][Eq.~109]{2017MNRAS.472.4204M}:
\begin{equation}
  \label{eq:5}
  F_y^{[0]}=0.644\frac{x_p^0}{\lambda}F_d,
\end{equation}
where
\begin{equation}
  \label{eq:6}
  F_d=\frac{\gamma(\gamma-1)GM(L-L_c)}{2\chi c_s^2}
\end{equation}
is the drag ($L < L_c$) or thrust ($L>L_c$) that the planet would
experience in circumstances in which the shear would be negligible
\citep{2019MNRAS.483.4383V}, and $x_p^0=a-r_c$ is the offset between
the semi-major axis and the corotation radius $r_c$, set by the radial
pressure gradient of the gas. In most of the discussion that follows we
restrict ourselves to the case $L>L_c$ (i.e. the net thermal force is dominated
by the heating of the ambient gas).  The first order term in
eccentricity is \citep{2019MNRAS.485.5035F}:
\begin{equation}
  \label{eq:7}
  F_x^{[1]} = \frac{a}{\pi\lambda}F_d e\left[f_x^C\cos (\Omega_pt) + f_x^S\sin(\Omega_pt)\right]
\end{equation} 
and
\begin{equation}
  \label{eq:8}
  F_y^{[1]} = \frac{a}{\pi\lambda}F_d e\left[f_y^C\cos (\Omega_pt) + f_y^S\sin(\Omega_pt)\right],
\end{equation}
where the coefficients $f_{x,y}^{S,C}$ are given at Eqs.~(172--175) of \citet{2019MNRAS.485.5035F}.
The time averaged torque is:
\begin{equation}
  \label{eq:9}
  \langle\Gamma\rangle=\left\langle(a+x)(F_y^{[0]}+F_y^{[1]})-yF_x^{[1]}\right\rangle.
\end{equation}
Using Eqs.~\eqref{eq:4}, \eqref{eq:5}, \eqref{eq:7}, \eqref{eq:8}
and~\eqref{eq:9} we arrive at:
\begin{equation}
  \label{eq:10}
  \langle\Gamma\rangle = F_da\left[ 0.644\frac{x_p^0}{\lambda}-0.58\frac{ae^2}{\lambda}\right].  
\end{equation} 
We see from this equation that the torque decreases as the
eccentricity increases. Writing 
$x_p^0=a\eta h^2$, where $\eta = O(1)$ is a dimensionless measure of the
offset between corotation and orbit \citep{2017MNRAS.472.4204M}, we obtain:
\begin{equation}
  \label{eq:11}
  \langle\Gamma\rangle = F_da\frac{a}{\lambda}\left[ 0.644\eta h^2-0.58e^2\right].
\end{equation}
This expression vanishes for $e=1.05\eta^{1/2}h$. However, for
$\eta\lesssim 1$, this corresponds to $e\sim h$, beyond the domain of
validity of case (i), for which we made the assumption $ea <
\lambda$. We therefore now turn to case (ii). In that case, the shear
is negligible \citep{2017MNRAS.465.3175M,2017arXiv170401931E} and the
planet is subjected to the force $F_d$ defined above, directed along
its direction of motion with respect to the gas, as long as it remains
subsonic. When the velocity with respect to the gas is almost equal to or
larger than the sound speed, the thermal force varies as a function of
the velocity, and in particular decays sharply in the supersonic regime
\citep{2019MNRAS.483.4383V}. The planet's
velocity in the frame corotating with the guiding centre is:
\begin{equation}
  \label{eq:12}
  (\dot x,\dot y) = ea\Omega_p[\sin(\Omega_pt),2\cos(\Omega_p t)]
\end{equation}
while in that same frame the gas velocity is: 
\begin{equation}
  \label{eq:13}
  (v_x^{\rm gas},v_y^{\rm gas}) = \left[0,-\frac 32\Omega_p (x+x_p^0)\right]
\end{equation}
hence the planet's velocity with respect to the gas is: 
\begin{equation}
  \label{eq:14}
  (\dot x|_{\rm gas},\dot y|_{\rm gas}) =
  \Omega_p\left[ea\sin(\Omega_pt),\frac{ea}{2}\cos(\Omega_p t)+\frac 32x_p^0\right].
\end{equation}
From this expression we can infer the components of the unitary vector
$\mathbf{n}=(n_x,n_y)$
having same direction as this velocity vector. The thermal force on
the planet is then $F_d\mathbf{n}$, and the torque on the planet is:
\begin{equation}
  \label{eq:15}
  \Gamma = F_d[(a+x)n_y-yn_x].
\end{equation}
Upon time averaging, we obtain (see appendix~\ref{sec:aver-therm-torq}):
\begin{equation}
  \label{eq:16}
  \langle\Gamma\rangle = F_da \left(1.20\frac{x_p^0}{ea}-1.54e\right),
\end{equation}
this expression being valid only for $ea>\lambda$. Again, we see that
the average torque decreases as the eccentricity increases.
\begin{figure}
  \includegraphics[width=\columnwidth]{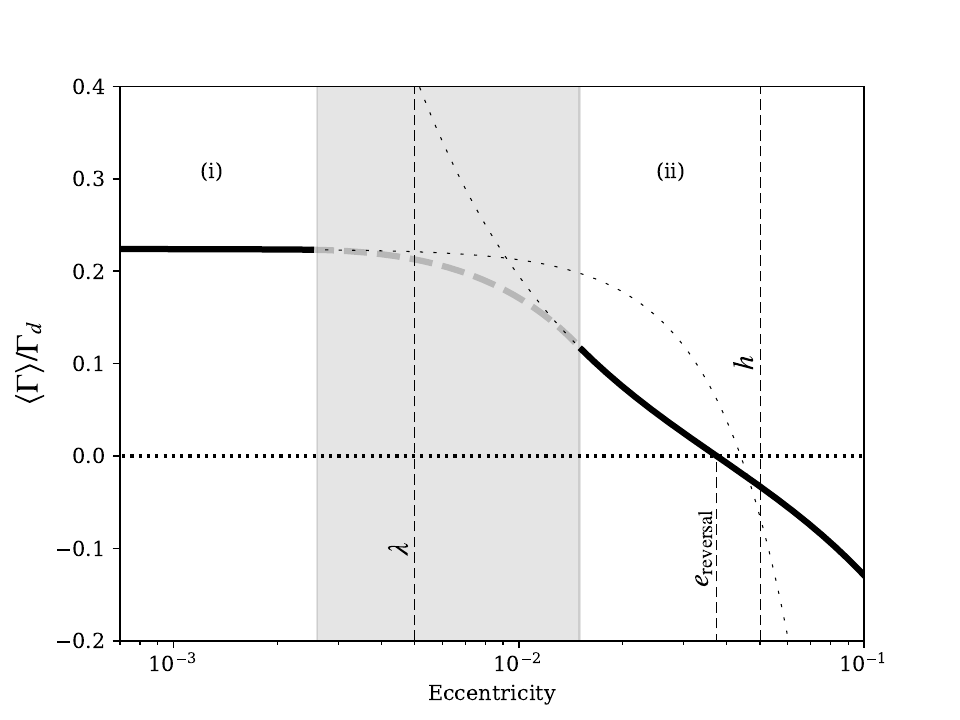}
  \caption{\label{fig:gvse} Time averaged torque as a function of
    eccentricity. The torque is normalised to $\Gamma_d=F_da$. The
    thick black solid lines show respectively regime~(i) (left)
    and regime~(ii) (right). From left to right, the vertical
    dashed lines show the thermal lengthscale, the eccentricity at
    which the torque changes sign and the aspect ratio. For this plot
    we have assumed $h=0.05$, $\lambda=5\cdot 10^{-3}a$ and
    $x_p^0=1.75\cdot 10^{-3}a$, corresponding to $\eta=0.7$. The grey
    band covers the eccentricity range for which neither regime is
    valid, the grey dashed line over this interval depicts what the
    transition between the two regimes may look like, and the thin
    dotted lines show the extrapolation of the two regimes beyond
    their domain of validity, using Eqs.~\eqref{eq:10}
    and~\eqref{eq:16}. We note that regime (ii) is valid as long as
    the planet is subsonic, so that for $e>h$ the actual curve would
    depart from the trend shown here.}
\end{figure}
We depict in Fig.~\ref{fig:gvse} the torque as a function of
eccentricity, for the two regimes~(i) and~(ii). As anticipated, the
torque remains positive over the domain of validity of regime~(i), while
it changes sign in regime~(ii) for
\begin{equation}
  \label{eq:17}
  e_\mathrm{reversal} = \sqrt{\frac{1.20}{1.54}\eta}h\approx
  0.9\eta^{1/2}h\lesssim h.
\end{equation}
Therefore, if the eccentricity of the planet, driven by the thermal
force, saturates to a value in excess of $e_\mathrm{reversal}$,
migration is inwards, at least if thermal forces dominate over
resonant forces. This migration reversal must be put solely on the
account of thermal forces. If, in addition, one takes into
  account the Lindblad and corotation torques, usually negative, the
  threshold eccentricity is smaller than this value, marginally
  so for very low mass planets, for which resonant torques are small
  compared to thermal torques.

\begin{figure}
  \centering 
  \includegraphics[trim={0cm 1.5cm 0 1.5cm},clip,width=1.2\columnwidth]{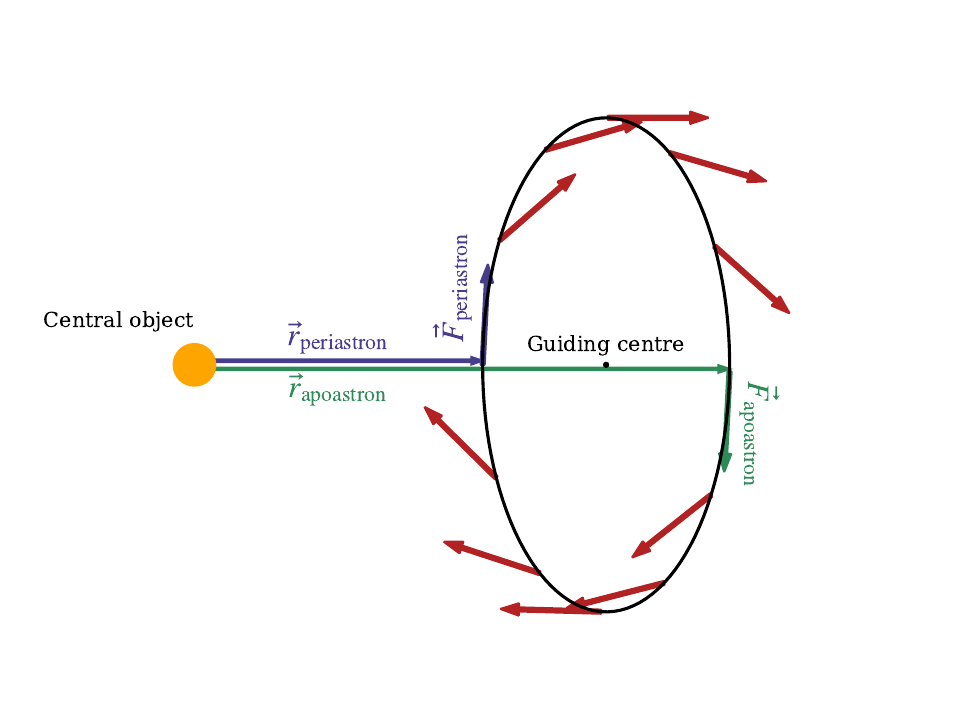}
  \caption{Depiction of the heating force along the epicycle of an
    eccentric planet. Its magnitude is a constant. In an inertial
    frame, rotation would be counter-clockwise, hence at periastron
    the torque of the heating force is positive, and negative at
    apoastron. The larger lever arm at apoastron entails a larger
    absolute value for the torque than at periastron.}
  \label{fig:croquis1}
\end{figure}

The reason for this reversal is depicted in Fig.~\ref{fig:croquis1},
which shows the heating force at several positions along the epicycle,
for a case without a corotation offset. This force has a constant
magnitude, and a varying orientation over an orbital period. Its lever
arm being larger at apoastron, the torque on the outer part of the
epicycle, which is negative, dominates.

This suggests that if the thermal force had a variable magnitude along
the epicycle, and a larger value on the inner part, the average torque
could remain positive. From Eq.~\eqref{eq:6}, we see that this may
happen if the luminosity, the thermal diffusivity or the sound speed
vary along the epicycle. For the time being, we remain agnostic of the
reasons of the variation, and simply assume that the magnitude of the
thermal force has a linear dependence on the distance to the central object:
\begin{equation}
  \label{eq:18}
  F_d(x) = F_d(0)\left(1+s\frac xa\right),
\end{equation}
where $s$ is the dimensionless slope of this dependence.
Repeating the calculation of the time average of the thermal force
with this new force expression (see Appendix~\ref{sec:aver-therm-torq-1}), we obtain, to
first order in $e$:
 \begin{equation}
   \label{eq:19}
   \langle\Gamma\rangle=F_da\left[1.20\frac{x_p^0}{ae} -(1.54+0.40s)e\right],
 \end{equation}
 where here $F_d$ stands for $F_d(0)$.  As anticipated, the torque
 remains positive for any value of the eccentricity if
 $s < s_c=-1.54/0.40\approx -3.85$. For such a value of the slope, the
 planet experiences a significantly larger thermal force at periastron
 than at apoastron. Consider a planet with an eccentricity driven to a
 value comparable to the aspect ratio: $e\sim h \sim 0.05$. The ratio
 of the force between periastron and apoastron is then
 $(1-s_ce)/(1+s_ce)\sim 1.5$.

 As said above, the magnitude of the heating force may vary because
 the thermal diffusivity, the sound speed or the luminosity of the
 planet varies. The first two quantities are intrinsic to the disc,
 and while they might exhibit such large variations over the
 relatively narrow range $[a(1-e),a(1+e)]$, the luminosity may vary
 much more, especially in the vicinity of dust rings.  Such rings
   can have widths comparable to or marginally smaller than the
   pressure lengthscale \citep{2018ApJ...869L..46D}, so that a planet
   having an eccentricity comparable to the disc's aspect ratio and
   its periastron in the ring will accrete significantly more there
   than at apoastron. In consequence it should easily fulfil
   the requirement that the luminosity at periastron be 50\% larger
   than that at apoastron and hence be subjected to a positive thermal
   torque, on average, provided a sizeable fraction of the
   potential energy of the accreted material is released as heat into
   the nearby gas on a timescale much shorter than the orbital
   timescale. In the following we will use extensive numerical
 simulations to check this hypothesis and identify the main properties
 of the dynamics of an accreting low-mass planet in the vicinity of a
 dusty ring.

 \begin{figure}
   \centering 
   \includegraphics[width=\columnwidth]{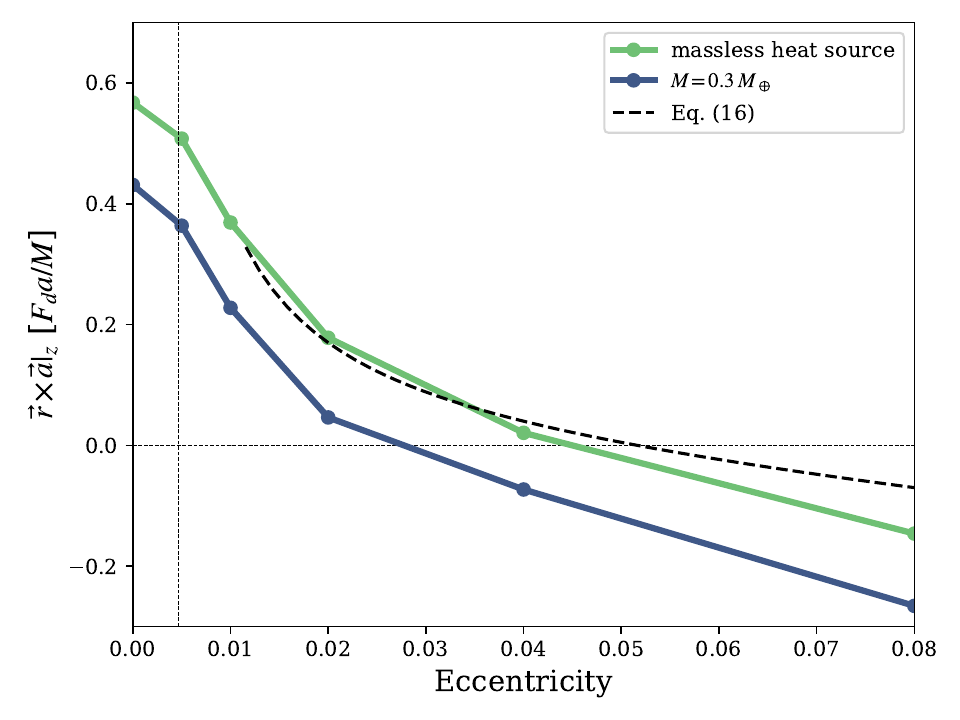}
   \caption{Specific torque as a function of eccentricity for a
     low-mass planet as described in the text, and thermal torque with
   same net luminosity for a massless perturber. The vertical dotted
   line shows the value of $\lambda/a$. The dependence expected from
   Eq.~(16) is shown starting at $e=2.5\lambda/a$. As the specific
   torque is undetermined for a massless object, we simply compute
   $\vec r\times \vec a$, where $\vec a$ is the acceleration imparted
   by the gas disturbance at the location of the planet.}
   \label{fig:massless}
 \end{figure}
 
 Before that, a few final remarks are in
 order. \citet{2023MNRAS.520.3286P} already noticed that it is
 unlikely that the torque reversal at larger eccentricity can be put on the
 account of the quenching of the corotation torque (since the reversal
 also happens when the corotation torque is substantially saturated),
 and interpret the decrease of the thermal torque with eccentricity as
 a cut-off. They provide an empirical fit with an exponential decay as
 a function of eccentricity (their Eq.~36). From the above we argue
 that the decrease of the thermal torque is rather a reversal, and
 that it may masquerade as an exponential cut-off only over a limited
 range of eccentricity, and only when the planet has a luminosity
 constant in time. Figure~\ref{fig:massless} shows the total specific
 torque exerted on a planet with mass $10^{-6}M_\star$ with a constant
 luminosity $L=5L_c$, in a disc with $\Sigma \propto r^{-2}$ (so that
 the total linear corotation, albeit positive\footnote{This linear
   corotation torque is the sum of three components, related
   respectively to the vortensity gradient, the temperature gradient
   and the entropy gradient \citep{2017MNRAS.471.4917J}.}, has a very
 small value: its quenching should have a negligible impact on the net
 torque). We see that the torque decays and changes sign for
 $e\lesssim h$ (the large corotation offset of our set of parameters
 requires a significant eccentricity to revert the torque). The top
 curve on that figure also shows the specific torque exerted on a
 vanishingly small mass (so that there is neither resonant nor cold
 thermal torque) with same net luminosity $5L_c$. That curve is very
 similar to the previous one, which clearly indicates that the net
 torque is dominated by the thermal torque, the offset between the two
 being attributable to the (subdominant) resonant and cold thermal
 torques. We also see that the thermal torque does not tend toward
 zero, but changes sign at higher eccentricity, as expected from the
 calculations above.

 The reversal of the time averaged thermal torque is similar to an
 effect found by \citet{2011ApJ...737...37M} for planets with an
 eccentricity significantly larger than the aspect ratio. The planet
 is then subjected to the dynamical friction of the gas, aligned with
 the relative velocity between the planet and the gas, much as the
 heating force, but with a negative sign. They find the
 dynamical friction to scale with the local midplane density of the
 gas. When the latter is constant (or increases outwards), the average
 torque is positive (their Fig.~10\footnote{That figure shows the
   timescale for the variations of $a$, which increases for $\alpha
   \leq 0$. At the same
   time $e$ decreases, so that the net torque is positive: both
   effects indeed add up to increase the area of the orbit.}).

 Our discussion has assumed a thermal torque dominated by heating
 effects. When the latter are absent, only the ``cold finger effect''
 \citep{2014MNRAS.440..683L} is at play and all our conclusions are
 reverted: the average torque, negative at small eccentricity, becomes
 positive above a certain value of the eccentricity. This is of
 limited interest though, as a finite eccentricity is usually not an
 equilibrium one for a planet experiencing only the cold and resonant
 torques. \citet{2023MNRAS.524.2705C} find indeed in these
 circumstances that the average power on the planet is negative,
 indicating an inward migration. The increase of the area of the
 orbit, required by the positive torque, is then achieved by lowering
 its eccentricity (as expected), rather than expanding its semi-major
 axis.

 \section{Numerical simulations}
 \label{sec:numer-simul}
 We make use of \texttt{FARGO3D} \citep{2019ApJS..241...25B} to perform three-dimensional numerical simulations of low-mass
 planets embedded in a gaseous protoplanetary disc, accreting dust
 from a narrow ring, and injecting the energy released by accretion
 into the nearby gas. The dust is modelled as a pressureless fluid \citep{2019ApJS..241...25B},
 and the ring is realised through a prior, two-dimensional simulation
 in the meridional plane allowing the dust to accumulate near a local
 pressure maximum in the gas. The accretion method for the dust
 requires a large resolution near the planet, so we use nested meshes
 centred on the average radial location of the planet \citep{2019MNRAS.483.4383V,2022MNRAS.509.5622V}. 

 Owing to the large computational cost of simulating one orbital
 period, we do not simulate the evolution of a growing planet over a
 large number of orbits. Rather, we perform many short term
 simulations that allow us to determine the variation rate of the
 semi-major and eccentricity as a function of the location and
 eccentricity of the planet, from which we determine, by
 interpolation, the orbital evolution of a low-mass planet over large
 amounts of time.

 \subsection{Governing equations}
 \label{sec:governing-equations}
 The equations that govern the evolution of the gas and dust are the
 continuity equation, the momentum equation, and an energy equation for
 the gas.
 \begin{eqnarray}
   \label{eq:20}
   \partial_t\rho_g+\mathbf{\nabla}\cdot(\rho_g\mathbf{v}_g)&=&0,\\
   \label{eq:21}
   \partial_t(\rho_g\mathbf{v}_g)+\mathbf{\nabla}\cdot(\rho_g\mathbf{v}_g\otimes\mathbf{v}_g)&=&-\mathbf{\nabla}P-\mathbf{\nabla}\cdot\mathbf{\tau}-\rho_g\mathbf{\nabla}\Phi,\\
   \label{eq:22}
   \partial_te_g+\mathbf{\nabla}\cdot(e_g\mathbf{v}_g)&=&-P \mathbf{\nabla}\cdot\mathbf{v}_g-\mathbf{\nabla}\cdot\mathbf{F}_h+Q,\\
   \label{eq:23}
   \partial_t\rho_d+\mathbf{\nabla}\cdot(\rho_d\mathbf{v}_d+\mathbf{j}_d)&=&0,\\
   \label{eq:24}
   \partial_t(\rho_d\mathbf{v}_d)+\mathbf{\nabla}\cdot(\rho_d\mathbf{v}_d\otimes\mathbf{v}_d)&=&-\rho_d\mathbf{\nabla}\Phi+k
   (\mathbf{v}_g-\mathbf{v}_d),
 \end{eqnarray}
 where $\rho_{g(d)}$ and $\mathbf{v}_{g(d)}$ denote the density and
 velocity of the gas (dust), $\tau$ the viscous stress tensor, $e_g$
 and $P$ the density of internal energy and pressure of the gas, $\Phi$
 the gravitational potential, $Q$ the heat source term arising from
 the release of accretional energy, $\mathbf{F}_h$ the heat flux
 arising from thermal diffusion. The expression of the viscous stress
 tensor has been given elsewhere \citep[e.g.][]{2016ApJS..223...11B}
 and is not repeated here. The kinematic viscosity $\nu$ that sets the
 magnitude of this tensor is evaluated using the $\alpha$ formalism
 \citep{ss73}: $\nu=\alpha_\nu \gamma r^2h^2\Omega_\mathrm{K}$. The pressure is linked to the density of
 internal energy through:
 \begin{eqnarray}
   \label{eq:25}
   P = (\gamma-1)e_g,
 \end{eqnarray}
 the gravitational potential is given by:
 \begin{eqnarray}
   \label{eq:26}
   \Phi(\mathbf{r}) = -\frac{GM_\star}{r}-\frac{GM}{[|\mathbf{r}-\mathbf{r}_p|^2+b^2]^{1/2}},
 \end{eqnarray}
 where $\mathbf{r}_p$ is the location of the planet and $b$
 the customary softening length of the potential.
 We do not include in this expression the indirect term arising from
 the acceleration of the star imparted by the planet because we expect this term to be negligible for the effect considered in this work. The heat source term
 arising from the accretion luminosity is:
 \begin{eqnarray}
   \label{eq:27}
   Q=L\delta(\mathbf{r}-\mathbf{r}_p),
 \end{eqnarray}
 where here only $\delta$ stands for Dirac's delta function,
 while the heat flux is given:
 \begin{eqnarray}
   \label{eq:28}
   \mathbf{F}_h = \chi\rho\mathbf{\nabla}\left(\frac{e}{\rho}\right),
 \end{eqnarray}
 where the thermal diffusivity $\chi$ is given by \citep[e.g.][]{2017MNRAS.471.4917J}:
 \begin{eqnarray}
   \label{eq:29}
   \chi = \frac{16(\gamma-1)\sigma T^3}{3\rho^2({{\cal R}/\mu})\kappa},
 \end{eqnarray}
where $\sigma$ is Stefan's constant, ${\cal R}$ the constant of ideal
gases, $\mu$ the mean molecular weight
and $\kappa$ the opacity, evaluated using the prescription of
\citet{1994ApJ...427..987B}. As we consider that the small grains, mainly responsible
  for the opacity, are well coupled to the gas, we do not take into
  account the increase of the dust-to-gas ratio in the ring for the
  evaluation of the opacity, and rely
  solely on the gas density to evaluate that quantity, using standard
  well-mixed gas and dust components with a dust-to-gas ratio $0.01$.
The additional flux $\mathbf{j}_d$ that features in the equation of continuity
on the dust captures the dust diffusion using the method given by
\citet{2018ApJ...854..153W}:
\begin{equation}
  \label{eq:30}
 \mathbf{j}_d=-\delta r^2h^2\Omega_K(\rho_g+\rho_g)\mathbf{\nabla}\left(\frac{\rho_d}{\rho_d+\rho}\right),
\end{equation}
where $\delta$ is a dimensionless coefficient that quantifies the
diffusion of dust. In all our calculations, we take
$\delta = \alpha_\nu = 10^{-4}$, in line with upper limits of
  turbulence obtained with ALMA \citep{2020ApJ...895..109F}. Finally,
the gas-dust friction coefficient $k$ that features in the momentum
equation on the dust is:
\begin{equation}
  \label{eq:31}
  k = \frac{\rho_d\Omega_\mathrm{K}}{\tau_s},
\end{equation}
where $\tau_s$ is the dimensionless stopping time (or Stokes number) of
the dust. We note that there is no equivalent term in the momentum
equation of the gas: we neglect the feed-back of the dust onto the
gas. 
In all the calculations presented here, we have adopted
$\tau_s=0.01$. This value is rather on the low side of the range
  of values for which a narrow ring forms. As the Stokes number
  increases, the width of the ring decreases, enhancing the variation
  of the accretion rate between periastron and apoastron, while the
  accretion radius of pebbles increases. It should therefore be kept
  in mind that the mechanism presented here could be even more
  efficient if the dust had a Stokes number higher than $0.01$.

\subsection{Initial conditions}
\label{sec:initial-conditions}
We specify hereafter the different fields of hydrodynamics quantities
at $t=0$, prior to the two-dimensional relaxation run.

\begin{align}
    \rho_g & =
             \frac{\Sigma_0}{\sqrt{2\pi}r_0h}\left(\frac{r}{r_0}\right)^{-\alpha-1}
             \sin(\theta)^{-\beta - \alpha-1 + \frac{1}{h^2}} \nonumber\\
  \label{eq:32}
           &\times\left[1+4\exp{\left(-\frac{(r-r_0)^2}{2r_0^2h^2}\right)}\right]\\
  \label{eq:33}
  e_g &= \frac{\rho_g h^2 G M_{\star}}{r(\gamma-1)}\\
  \label{eq:34}
  v^\phi_g&=\sqrt{\frac{GM_\star}{r}}\left[1-(\beta+\alpha+1)h^2\right]^{1/2}-\Omega_\mathrm{frame}r\sin\theta\\
  \label{eq:35}
  v^\phi_d&=\sqrt{\frac{GM_\star}{r}}-\Omega_\mathrm{frame}r\sin\theta
\end{align}
The dependence of $\rho_g$ on the colatitude $\theta$ given in
  Eq.~\eqref{eq:32} is that worked out by \citet{2016ApJ...817...19M}
  in locally isothermal discs in which the sound speed is a function
  of the spherical radius. It tends towards the standard Gaussian
  profile near the midplane ($\theta\approx\pi/2$).  In
Eqs.~\eqref{eq:32}--\eqref{eq:35}, $r_0$ is a reference radius close
to the ring maximum, $\Sigma_0$ would be the disc's surface density at
$r_0$ in the absence of the ``bump'' imposed by the last factor of
Eq.~\eqref{eq:32}, $h$ is the aspect ratio of the gaseous disc
(assumed constant), $\alpha$ is the ``slope'' of surface density
($\alpha=-d\log\Sigma/d\log r)$ and $\beta$ that of temperature
($\beta=-d\log T/d\log r$). We choose $h=0.05$,
$\Sigma_0 = 10^{-3}M_\star/r_0^2$, $\alpha=1/2$, and have $\beta=1$ as
per our assumption of a constant aspect ratio. We specify to the
  case $r_0=10$~au and $M_\star$ is a solar mass. The value of the
  surface density then amounts to $1.65$ times that of the MMSN at
  $10$~au. The slope of surface density corresponds to a shallow decay
  routinely used in simulations with smooth discs. We anticipate that
  our results are hardly impacted, if at all, by this value, as most
  of the effect we present here arises from dominant thermal forces in
  the ring, and since the variation of the surface density over the
  radial excursion of the planet is small.

The initialisation of the dust density is done in a way that handles
poorly resolved layers in the vertical direction. Indeed, when
  the disc is resolved over a very small number of zones in the
  vertical direction, sampling the density using the standard Gaussian
  expression evaluated at the zone centres does not guarantee that a
  vertical integration yields the desired surface density. Rather, we
  evaluate the density as the (discrete) derivative of an error
  function, which by construction allows to enforce the value of the
  integral of that quantity (i.e., the surface density). When the
  resolution is sufficient, the result is indistinguishable from the
  direct evaluation of a Gaussian profile:
\begin{equation}
  \label{eq:36}
    \rho_d= \Sigma_0Z\left(\frac{r}{r_0}\right)^{-\alpha}
    \!\left[\text{erf}\left(\frac{\pi/2-\theta_-}{\sqrt 2 h_d}\right)-\text{erf}\left(\frac{\pi/2-\theta_+}{\sqrt
          2 h_d}\right)\right] \!\frac{1}{2r\Delta\theta},\\
  \end{equation}
  where $Z$ it the dust-to-gas ratio,
  $h_d=\sqrt{\delta/(\tau_s+\delta)}h$ \citep{2007Icar..192..588Y} is
  the aspect ratio  of the dusty disc, $\theta_-$ ($\theta_+$) is the
  lower (upper) bound of a zone in colatitude, and
  $\Delta\theta\equiv\theta_+-\theta_-$. Regardless of the value of
  $h_d/\Delta\theta$, the formulation of Eq.~\eqref{eq:36} always
  yields a column density of dust $Z\Sigma_0(r/r_0)^{-\alpha}$. While
  the dust layer is correctly resolved on the patch of highest resolution,
  this may not be true on the base mesh, and this treatment avoids to
  have discontinuities of the column density of dust across
  mesh boundaries. In all simulations presented here we
  have  $Z=0.01$.

\subsection{Meshes and setup}
\label{sec:meshes-setup}
The disc is described on a set of four nested, spherical meshes. The
azimuthal extent of the base mesh is $1.6$~rad, largely smaller than
$2\pi$, to save computational time. The frame corotates with the
planet, so that its angular frame oscillates in time when the planet
is eccentric. This entails that the planet's trajectory is a radial
segment, and allows to restrict the azimuthal extent of the refined
patches. The extent of the different levels is specified in
Tab.~\ref{tab:mehses}.

\subsection{Relaxation of initial conditions: creation of the dusty ring}
\label{sec:ring-dust:-2d}
The initial conditions outlined in
section~\ref{sec:initial-conditions} do not correspond to a disc in
rotational equilibrium: among other things, the bump introduced in the
density (and therefore pressure) is not reflected in the azimuthal
speed. Besides, the dust density is simply a scaled version of that of
the gas: the dust has not yet accumulated in the vicinity of the pressure
maximum, so that there is no thin dusty ring in the initial
conditions. For these reasons, we perform a two-dimensional
calculation in the $(r,\theta)$ plane in order to allow the disc to
relax toward rotational equilibrium, and to allow the dust ring to
form. This initial relaxation is performed only on the base mesh
($\ell=0$, see Tab.~\ref{tab:mehses}), with only one zone in azimuth.

\begin{figure}
  \centering
    \includegraphics[width=\columnwidth]{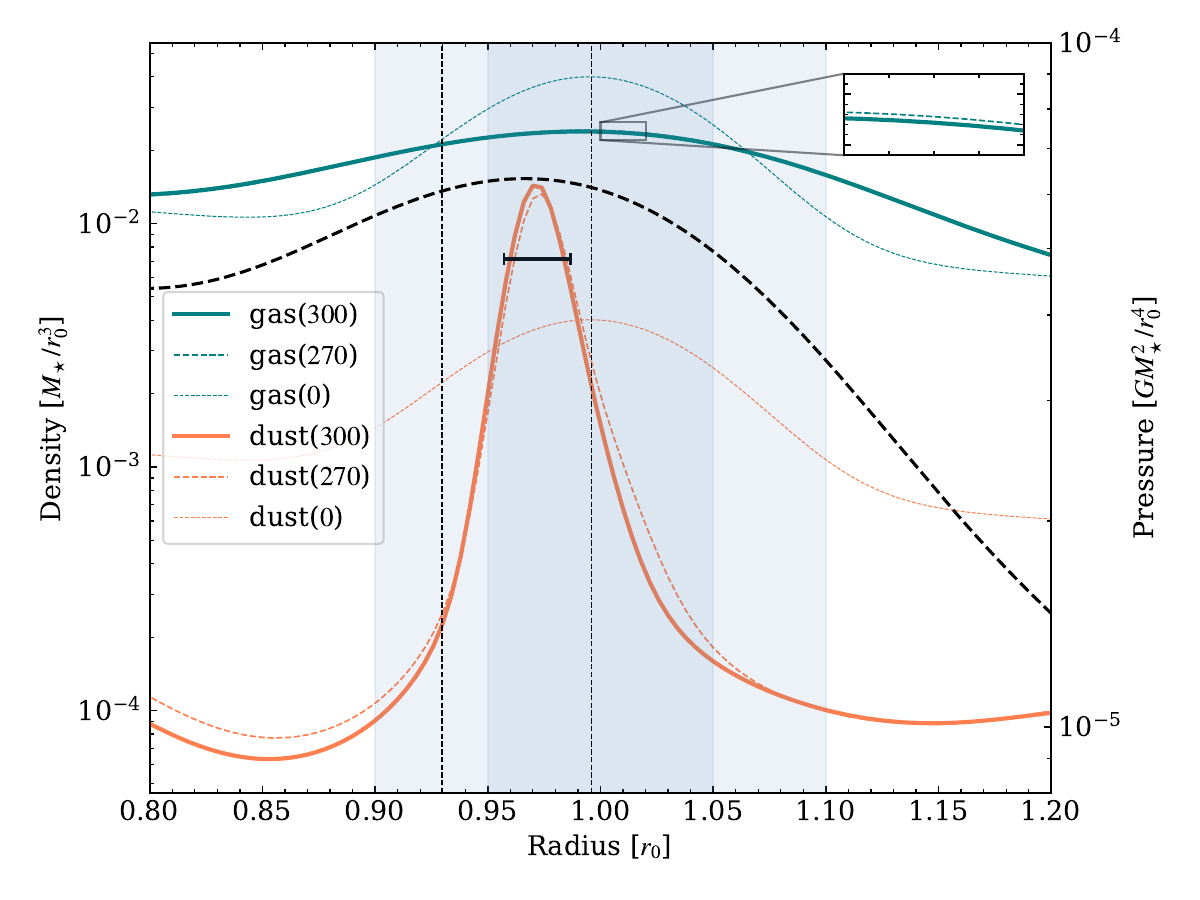}
    \caption{Midplane densities (solid lines) and pressure (dashed
      line) after 300 orbits. The thin dotted lines show the initial
      profiles of gas and dust densities, while the thicker dotted
      lines show these profiles after 270 orbits. The scale of pressure (right
    vertical axis) is chosen so as to show that the thin dust ring has
  its peak density at the pressure maximum. The horizontal segment on the peak
  of dust density shows the FWHM expected for the dust ring. The radial domain
  covers what would be the radial extent of the mesh at level
  $\ell=1$ (for a planet with $a=1$), while the light vertical bands show the extent of levels $\ell=2$
  and $\ell=3$. These bands are shown for informational purpose only,
  as nested meshes are not used during the relaxation run. For
    future reference, the
    vertical dashed lines show the location of the two migration traps
    of a cold, low mass planet on a circular orbit, subjected to the
    gas and dust torques, discussed later in the text.}
    \label{fig:ring}
  \end{figure}

We find that $270$~orbital periods at $r_0$ are enough to obtain
profiles that do not vary significantly in time and allow us to
measure the effect of the dust ring on an eccentric planet. The
initial bump of gas density and pressure widens significantly, and its
amplitude decays, but a local maximum of pressure remains, which
allows the dust to converge and form a thin ring. We
show the converged profiles in Fig.~\ref{fig:ring}. A Gaussian fit
near the pressure maximum upon relaxation yields a width of the
pressure bump $w_g=2.5r_0h$ and a relative amplitude $\sim 2$, which
indicates that it should not be subjected to the Rossby Wave
Instability (RWI) and that the dust trap should survive \citep[][their
Fig.~2]{2023ApJ...946L...1C}.  The mass of the dust ring obtained
after relaxation is $10$~M$_\oplus$. From the width of the pressure bump,
one can estimate the expected width of the dust ring
\citep{2018ApJ...869L..46D}:
  \begin{equation}
    \label{eq:37}
    w_d=w_g\sqrt{\alpha_\nu/\tau_s}.
  \end{equation}
  The corresponding FWHM is shown on Fig.~\ref{fig:ring} and matches
  satisfactorily the width of the dust ring.  While, as stated above,
  the pressure bump is not expected to experience the RWI, it could
  still be subjected to a more complex version of that instability
  involving dust \citep[DRWI,][]{2023MNRAS.526...80L}, in particular
  that involving a mild pressure bump. Since we do not take
  into account the dust feedback onto the gas, this instability should
  not happen in our setup. Furthermore, the short timescale over which
  our three-dimensional runs are performed would preclude the
  appearance of this instability.

\subsection{Restart as a three-dimensional run}
\label{sec:restart-as-three}
Once the ring of dust has formed and the disc profiles have converged
to profiles constant in time, we extend the domain for $\phi$ over the
range $[-.8,.8]$ with $N_\phi=400$ (hence
$\Delta_\phi=4\times10^{-3}$). We also include three levels of
refinement in order to reach $\Delta_{r,\phi,\theta}=5\times10^{-4}$,
using the patches detailed in Tab.~\ref{tab:mehses}. We insert a
low-mass planet on a non-inclined, eccentric orbit, using a sinusoidal
taper over a timescale that corresponds to one orbital period at
$r_0$:
\begin{equation}
  \label{eq:38}
  M(t)=
  \begin{cases}
      M\sin^2\left(\frac{\Omega_0t}{4}\right) & \text{if $t<2\pi/\Omega_0$}\\
      M & \text{otherwise,}
    \end{cases}       
  \end{equation}
  where $\Omega_0=(GM_\star/r_0^3)^{1/2}$ is the Keplerian frequency
  at $r_0$. This tapering is immaterial for the problem at hand: the
  timescale of the disc's response is significantly shorter than the
  orbital period \citep{2017MNRAS.465.3175M, 2017arXiv170401931E},
  hence the variation rates of the semi-major axis and eccentricity,
  measured over the last orbital period, are virtually independent of
  the mass growth imposed over the first orbit.  The planet accretes
  dust from the disc and releases the energy obtained from accretion
  into the ambient gas. We detail the implementation of these two
  processes below.

\begin{table*}
    \centering 
    \begin{tabular}{l|c|c|c|c}
        \hline 
        $\ell$ & $(N_\phi,N_r,N_\theta)$ & $(\phi_\text{min},\phi_\text{max})$ & $(r_\text{min},r_\text{max})$ & $(\theta_\text{min},\theta_\text{max})$\\
        \hline 
        $0$ & $(400,200,30)$  &  $[-.8,.8]$ &  $a+[-0.4,0.4]r_0$ & $[\frac{\pi}{2}-.12,\frac\pi 2]$\\
        $1$ & $(200,200,60)$  &  $[-.2,.2]$ &  $a+ [-0.2,0.2]r_0$ & $[\frac{\pi}{2}-.12,\frac\pi 2]$\\
        $2$ & $(200,200,60)$  &  $[-.1,.1]$ &  $a+[-0.1,0.1]r_0$ & $[\frac{\pi}{2}-.06,\frac\pi 2]$\\
        $3$ & $(200,200,60)$  &  $[-.05,.05]$ &  $a+[-.05,.05]r_0$ &$[\frac{\pi}{2}-.03,\frac\pi 2]$\\
        \hline 
    \end{tabular}
    \caption{\label{tab:mehses}Size and extent of each level of the
      system of nested meshes. The central radius of each level
      corresponds to the semi-major axis of the planet.}
\end{table*}{}

\subsection{Dust accretion}
\label{sec:dust-accretion}
We model the accretion of dust by removing a fraction of the material
in the cells nearest to the planet. Our rationale for using this procedure is that the size of the cells on the layer of highest resolution is significantly smaller than the accretion radius within which pebbles are in the settling regime, even for the smallest planetary mass considered in this study. Furthermore, the settling time at the distance from the planet comparable to the resolution is shorter than or comparable to the timestep arising from the Courant condition. For the cells $(i,j,k)$ that comply with
$|\phi_p-\phi_{ijk}|<\Delta \phi$, $|r_p-r_{ijk}|<\Delta r$ and
$|\theta_p-\theta_{ijk}|<\Delta \theta$, i.e. the cells nearest to the planet, the
dust is removed using the following recipe. We evaluate:
\begin{equation}
  \label{eq:39}
    \Delta \rho_{ijk} = 
     \rho_{ijk} \left(1-\frac{\phi^{'}_{ijk}}{\Delta 
         \phi}\right)\left(1-\frac{r^{'}_{ijk}}{\Delta 
         r}\right)\left(1-\frac{\theta^{'}_{ijk}}{\Delta \theta}\right), 
\end{equation}
where $\xi^{'}_{ijk}=|\xi_p-\xi_{ijk}|$ for $\xi\equiv\phi,r,\theta$.\\
The new dust density is then:
\begin{equation}
  \label{eq:40}
    \rho'_{ijk}=\rho_{ijk}-\Delta \rho_{ijk}
\end{equation}
and the accretion rate is simply:
\begin{equation}
  \label{eq:41}
    \dot{M} = \sum_{ijk}\frac{\Delta\rho_{ijk}V_{ijk}}{ \Delta t }, 
  \end{equation}
  where $V_{ijk}$ is the volume of the cell $(i,j,k)$ and the sum is
  performed over the cells for which the density was changed.

  In order to validate this approach, we have compared the accretion rates obtained with this
    procedure to those obtained semi-analytically by evaluating the
    accretion radius as a function of the relative velocity between the
    planet and the pebbles. We find a reasonable agreement between the two,
    generally within a factor of two. As we shall see in
    section~\ref{sec:vari-from-fiduc}, we find that the accretion efficiency
    depends on the planet's luminosity. Should we had forced the
    simulations to use an accretion rate based on formulae that ignore
    the planet's luminosity, we would have missed this side result.

The accreted mass of the pebbles is not added to that of the
  planet. Likewise, the momentum accreted is not added to the
  planet. We shall discuss this in section~\ref{sec:other-limitations}.

\subsection{Heating}
The accretion rate $\dot{M}$ is then used to compute the luminosity of the planet:
\begin{equation}
  \label{eq:42}
    L = \frac{GM\dot{M}}{R_p}, 
\end{equation}
where $R_p=\left(\frac{3M}{4\pi\rho_p}\right)^{1/3}$ is the planet's
physical radius. In this expression we adopt for the density of the
planet $\rho_p=3$~g~cm$^{-3}$. Note that in Eq.~\eqref{eq:42} we use
$M$, the mass at the end of the short, progressive growth (see
section~\ref{sec:restart-as-three}) and the corresponding physical
radius $R_p$, rather than $M(t)$ and a corresponding instantaneous
radius $R_p(t)$. The luminosity is therefore overestimated over the
initial mass ramp.  However, this initial growth stage is entirely
artificial and has no impact on the variation of the orbital elements
measured on the last orbit, owing to the short response time of the
thermal force.

The luminosity is then used to increase the density of internal energy
of the gas in the neighbouring cells, i.e. those which fulfil the same
distance criteria as those used for the accretion of dust (see section
\ref{sec:dust-accretion}):
\begin{equation}
  \label{eq:43}
\Delta{e}_{ijk} = 
  L \left(1-\frac{\phi^{'}_{ijk}}{\Delta 
         \phi}\right)\left(1-\frac{r^{'}_{ijk}}{\Delta 
         r}\right)\left(1-\frac{\theta^{'}_{ijk}}{\Delta \theta}\right)\frac{\Delta t}{V_{ijk}}.
   \end{equation}
   This method is the same as that used by \citet{2017arXiv170401931E}.

 A more sophisticated prescription for the
     injection of heat into the nearby gas has been investigated by
     \citet{2020MNRAS.495.2063V}, and found to essentially yield same
     results as the simpler prescription considered here.  The reason
     for that is that the asymmetry of the heated region, which
     induces the thermal force, kicks in at distances from the planet
     larger than the release radius.

   \section{Results}
   \label{sec:results}
   We present hereafter the results of our numerical simulations. We
   firstly present in detail our fiducial calculation, that of a
   one-Earth mass protocore orbiting in  the vicinity of the ring
   obtained at section~\ref{sec:ring-dust:-2d}, before presenting
   additional results for other masses and/or a slightly modified
   setup.

   \subsection{Fiducial calculation}
   \label{sec:fiducial-calculation}
\begin{figure*}
     \includegraphics[width=\textwidth]{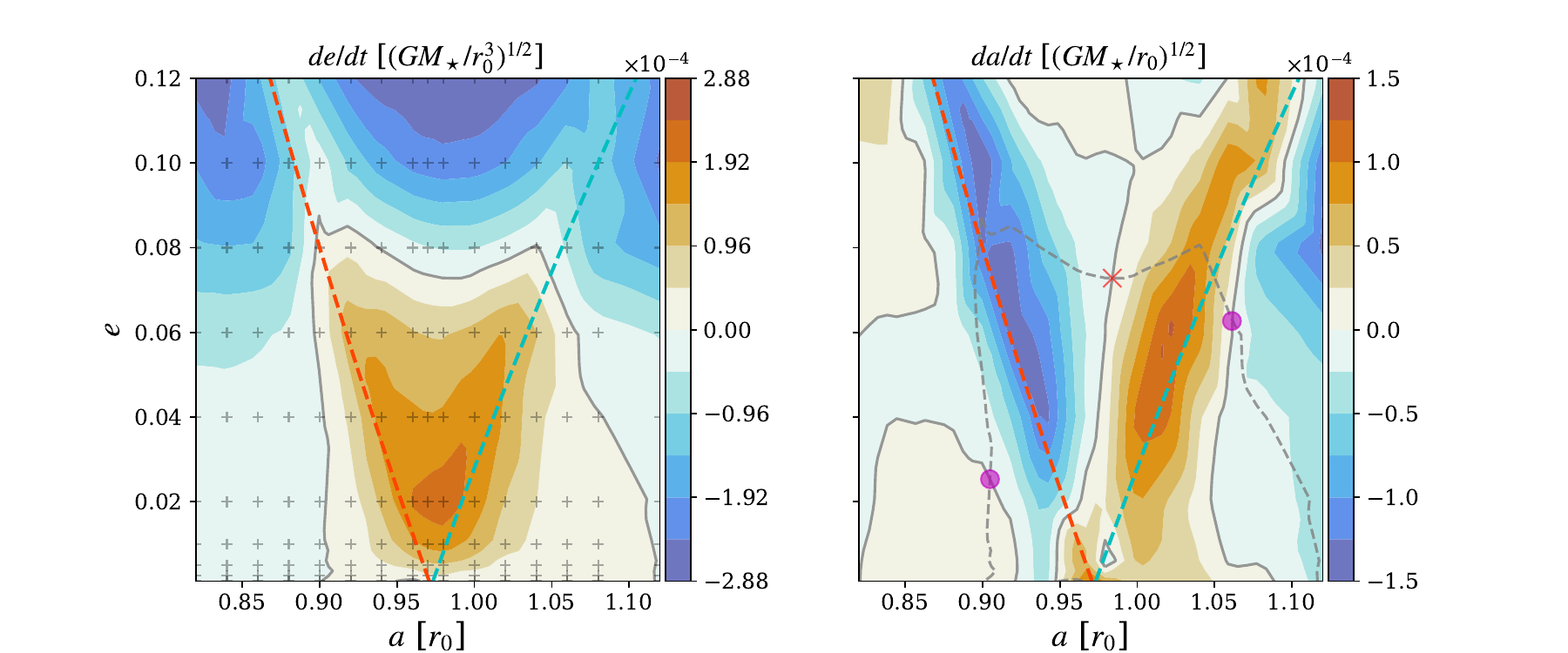}
     \caption{Time derivative of eccentricity (left) and semi-major
       axis (right) as a function of semi-major axis and eccentricity,
       for our fiducial exploration. The red (blue) inclined dashed
       line corresponds to planets that have their apoastron
       (periastron) at the maximum of the dust ring. Planets between
       these two lines cross the ring maximum. The crosses on the left
       plot correspond to the values used in our runs. The results are
       interpolated, upon triangulation, to produce the maps. For
       legibility reasons, the same crosses have not been reproduced
       on the right plot. The dashed grey contour of the right plot is
       a copy of the contour $\dot e=0$ from the left plot. Its
       intersection with the solid grey contours corresponding to
       $\dot a=0$ yields locations where the orbit has constant
       semi-major axis and eccentricity. Magenta filled circles denote
       stable equilibrium points while the cross identifies an
       unstable equilibrium (see text for details).}
     \label{fig:deda}
   \end{figure*}

   Our fiducial calculation consists of a systematic exploration of
   the semi-major axis, eccentricity space for a one Earth-mass
   planet. The set of semi-major axis values is
   $\{0.82r_0+0.02r_0i \mbox{~for~} i\in[0,13]\} \cup\{0.97r_0\}\cup
   \{1.12r_0\}$ (16 values in total), while the set of eccentricity
   values is
   $\{1.25\cdot 10^{-3}\times 2^i\mbox{~for~}i\in[0,3]\} \cup
   \{0.02i\mbox{~for~}i \in[1,6]\}$ (10 values in total). The total
   number of runs of this fiducial exploration is therefore
   $160$. Each run lasts three orbital periods at $r_0$.  The planet
   is free to move under the disc's force, which we evaluate by
   subtracting the azimuthal average of the density from each given
   cell prior to evaluating the force it exerts on the planet. This
   method has been suggested as a workaround for the spurious shift of
   resonances that occurs when the disc's self-gravity is discarded
   \citep{2008ApJ...678..483B,2016ApJ...826...13B,2020A&A...635A.204A}. While
   this effect would at best be minute here, it is also important to
   use this method when working on a wedge of the disc, as we do
   here: if the planet does not lie exactly on the bisector line of the wedge,
   it would be subjected to a strong, spurious force arising from the
   asymmetry of the mass distribution around it, even if the disc is unperturbed.
   We measure the drift of its semi-major axis and eccentricity over
   the last planetary orbit. Namely, if $t_f$ is the last time at
   which we get a measure of $a$ and $e$ in the log files, we seek the
   value of $a$ and $e$ at $t_i=t_f-2\pi\sqrt{a^3/GM_\star}$. Except
   for $a=r_0$, this time is in general not found in the log files
   (owing to our uniform time sampling across all runs) and the
   corresponding value of $a$ and $e$ is obtained by linear
   interpolation. The time derivatives of $a$ and $e$ are then
   approximated as:
   \begin{equation}
     \label{eq:44}
     \dot a \approx \frac{a(t_f)-a(t_i)}{t_f-t_i}\mbox{~~~and~~~}
     \dot e \approx \frac{e(t_f)-e(t_i)}{t_f-t_i}.
   \end{equation}
We show in Fig.~\ref{fig:deda} a map of the time derivative of
eccentricity and semi-major axis as function of these two
quantities. These maps show that
\begin{itemize}
\item the eccentricity is excited up to 
values slightly larger than the disc's aspect ratio for planets whose 
orbit cross the ring. When the orbit does not cross the ring, the 
eccentricity is damped (for planets inside the ring) or driven to 
moderate values (for planets on the outside);
\item the semi-major axis of planets that have their periastron in the
  ring tends to increase with time, whereas it decreases with time for
  planets that have their apoastron in the ring.
\end{itemize}
We identify the contours where the eccentricity or the semi-major axis
remain constant. Their intersection correspond to orbits that do not
evolve in time: they have $\dot e=0$ and $\dot a=0$. These orbits are
not necessarily stable. Consider for instance the point near
$a=0.98r_0$ and $e=0.073$, represented by a cross. This point is
obviously unstable as a planet displaced to the right (left), will
experience an increase (decrease) of its semi-major axis.

\begin{figure}
  \includegraphics[width=.9\columnwidth]{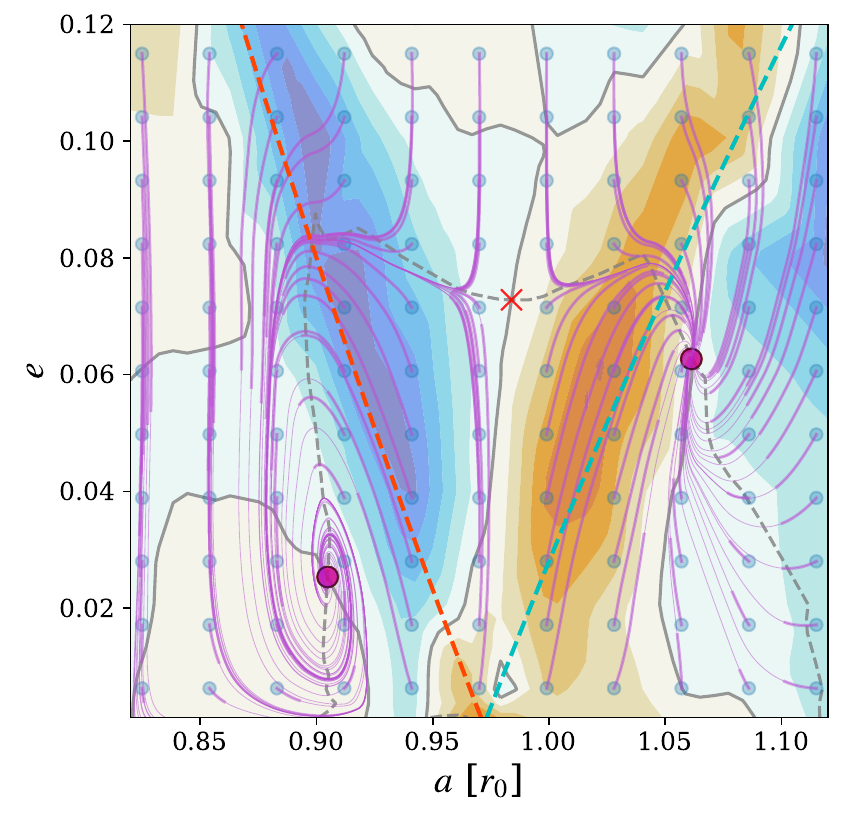}
  \caption{Trajectories in the $(a,e)$ plane for our fiducial
    exploration, superposed on the map of $\dot a$. The $11\times 11$
    blue dots are the starting points. Each trajectory is represented
    by a thicker line for an integration time up to
    $500\sqrt{r_0^3/GM_\star}$, i.e. $\approx 80$~orbital periods at
    $r_0$, or $\approx 2500$~yrs if the central object has a solar
    mass.}
  \label{fig:traj}
\end{figure}

\begin{figure}
  \includegraphics[width=\columnwidth]{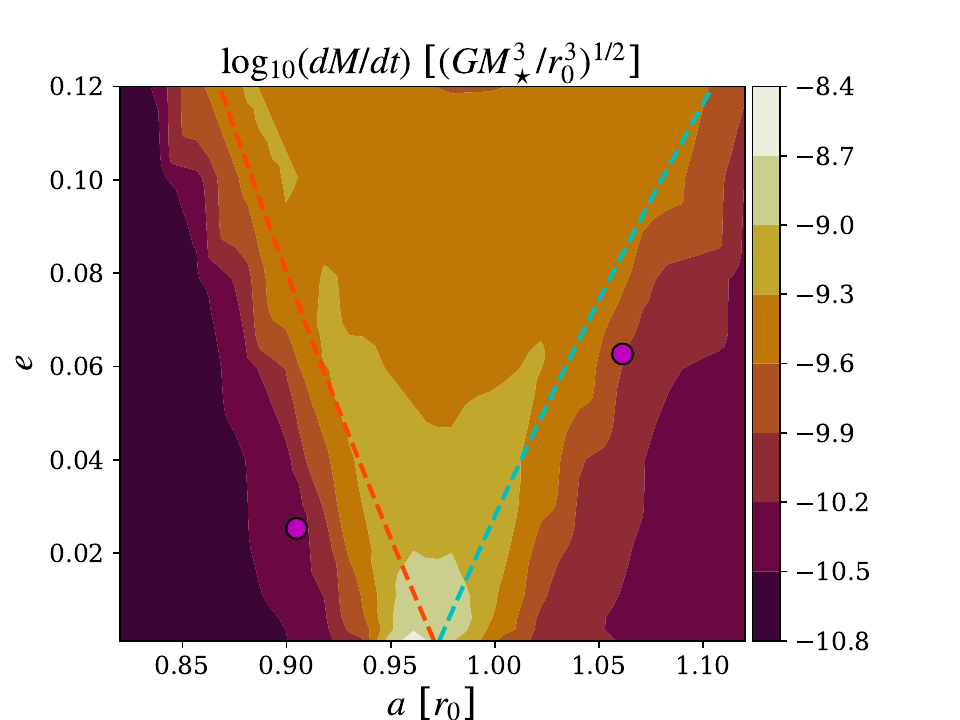}
  \caption{Orbital averaged mass accretion rate in our fiducial runs,
    as a function of semi-major axis and eccentricity. The two fixed
    points identified in Fig.~\ref{fig:deda} are marked as magenta discs.}
  \label{fig:dmdtfidu}
\end{figure}

   There are two other points, however, identified by a magenta disc on
   the right plot of Fig.~\ref{fig:deda}, that may correspond to
   stable orbits. We study the trajectories in the $(a,e)$ plane to
   assess this stability, using the
     interpolated maps of $\dot e$ and $\dot a$. The results are shown in
   Fig.~\ref{fig:traj}, which confirms that these two points do indeed
   correspond to stable orbits. Depending on its starting location in
   the $(a,e)$ plane, a planet will end up at one of these two points,
   over a timescale of order $10^3-10^4$~years. 
We evaluate the accretion rate at these fixed points. It is calculated
in the same manner as $\dot e$ and $\dot a$: we evaluate it over the
last orbit, and we interpolate it at the location of the fixed
points. We plot the map of $\dot M(a,e)$ in Fig.~\ref{fig:dmdtfidu},
and show the location of the two fixed points.
At the outer fixed point, we measure $M/\dot M\approx 22\cdot
10^3\Omega_p^{-1}$ or $1.1\cdot 10^5$~yrs, while at the inner fixed
point we measure: $M/\dot M\approx 58\cdot
10^3\Omega_p^{-1}$ or $2.9\cdot 10^5$~yrs.
\subsection{Interpretation of the results}
\label{sec:interpr-results}
\begin{figure*}
  \includegraphics[width=\textwidth]{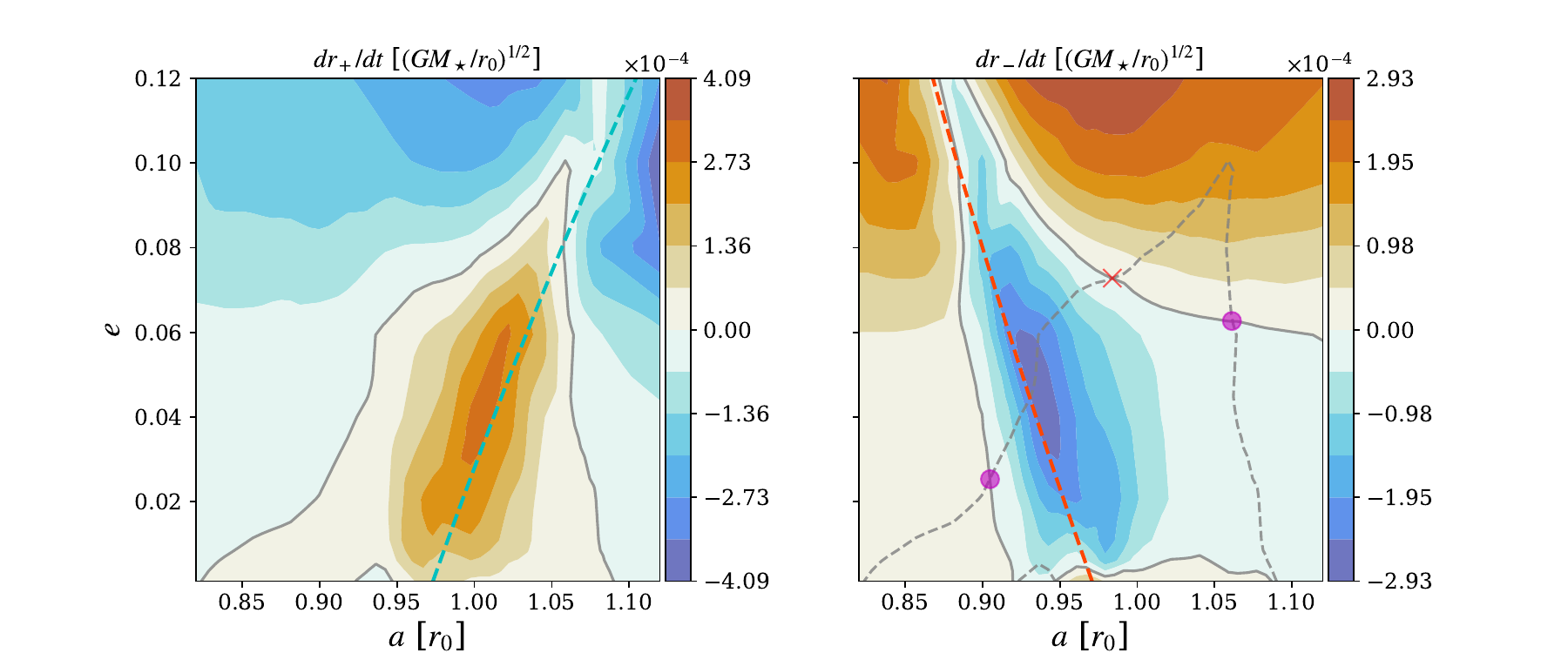}
  \caption{Time derivative of the apoastron (left) and of the 
    periastron (right). For an eccentricity smaller or of the order of 
    the aspect ratio, the main effect of the interaction with the ring 
    is an increase of the apoastron for planets that have 
    their periastron in the ring (those on the blue dashed line of 
    the left plot) and a decrease of the periastron for 
    planets that have their apoastron in the ring (those on the red 
    dashed line of the right plot). Blue contours correspond to the 
    value zero. The dashed grey contour of the right plot is a copy of the 
    contour $\dot r_+=0$ of the left plot. We recover the two fixed 
    points corresponding to $\dot a=0$ and $\dot e=0$.}
  \label{fig:periapo}
\end{figure*}

\begin{figure}
  \includegraphics[width=.5\columnwidth,trim={50mm 0 50mm 0},clip]{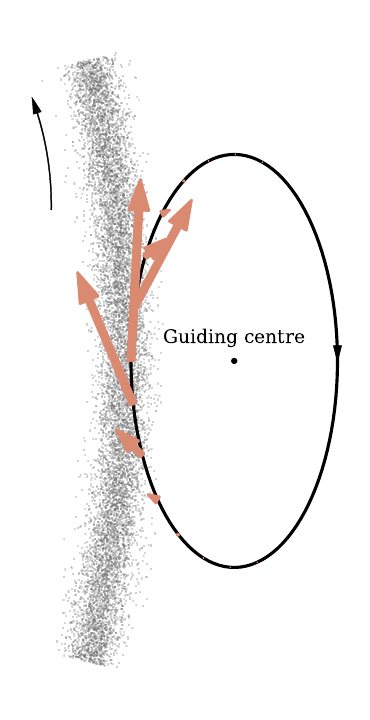}\includegraphics[width=.5\columnwidth,trim={50mm 0 50mm 0},clip]{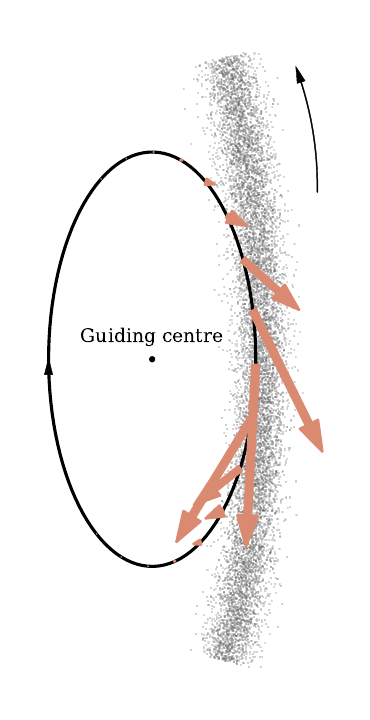}
  \caption{Sketch of a planet that has its periastron in the ring (left) 
  or its apoastron in the ring (right). The rotation of the ring in an 
inertial frame is counterclockwise, as indicated by the arrow near the 
top of the plots. In the frame corotating with the guiding centre, the 
epicycle is described in the clockwise direction. The left (right) plot 
corresponds to a planet on the blue (red) dashed line of the left (right) 
plot of Fig.~\ref{fig:periapo} }
  \label{fig:sketchring}
\end{figure}

Our results are compatible with the action of a strong thermal 
force that arises every time the planet goes into the ring.
They can be recast in terms of the time derivative of the
periastron and apoastron distances, which are respectively:
\begin{equation}
  \label{eq:45}
  r_-=a(1-e)\mbox{~~~and~~~}r_+=a(1+e).
\end{equation}
We see in Fig.~\ref{fig:periapo} that planets that have their
periastron in the ring experience a strong increase of their
apoastron, and nearly no variation of the periastron. Such planets are
subjected to a strong thermal force as they pass through periastron
(see left sketch of Fig.~\ref{fig:sketchring}). The net effect of this
passage is an increase of their orbital velocity at periastron. This
entails that the periastron distance is conserved, while their
apoastron is increased by some amount $\Delta r_+$ (and their
semi-major axis by $\Delta r_+/2$). Reciprocally, a planet that has
its apoastron in the ring is subjected to a strong, resistive thermal
force during its passage through apoastron (see right plot of
Fig.~\ref{fig:periapo} and right sketch of Fig.~\ref{fig:sketchring}),
resulting in a decay of the periastron distance (and of the semi-major
axis by half that of the periastron).

A planet with initially a semi-major axis larger than the ring's
radius will park itself at a semi-major axis and eccentricity such
that it accretes from the ring at periastron, and such that the kicks
of energy and angular momentum it receives there from the thermal
force make up for the decay of these quantities over the rest of their
epicycle. This corresponds to the outer fixed point. This implies
  that the trapping at the outer fixed point relies only on the
  existence of the narrow dusty ring, not on the existence of the
  pressure bump: the planet always remain in a region where the
  pressure decreases monotonically outwards. Should a ring exist in
  the absence of a pressure bump, we would expect a similar effect to
  operate.

The inner fixed point corresponds to an effect somewhat similar to the
case studied in section~\ref{sec:torq-revers-diff}: above a critical
eccentricity that is a fraction of the aspect ratio, migration
reverses from outward to inward, and at the critical eccentricity for
reversal, migration stops. Too far from the ring, the accretion
luminosity is not sufficient to maintain a finite eccentricity and the
latter decays, while too close to the ring, the eccentricity is driven
to values higher than the critical value for migration reversal. In
between these two cases, there is a distance to the ring such that the
eccentricity is driven precisely to the value for which migration
stalls, which corresponds to the inner fixed point.

\subsection{Variation from the fiducial calculation: efficiency of
  accretion}
\label{sec:vari-from-fiduc}
\begin{figure*}
  \includegraphics[width=.95\columnwidth]{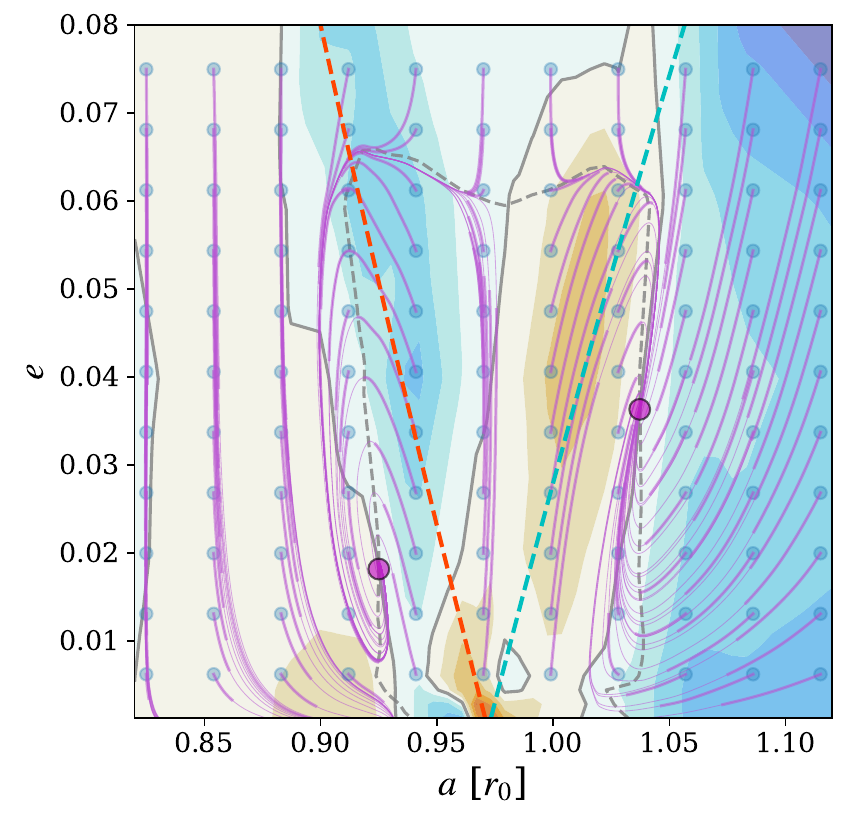}\includegraphics[width=.95\columnwidth]{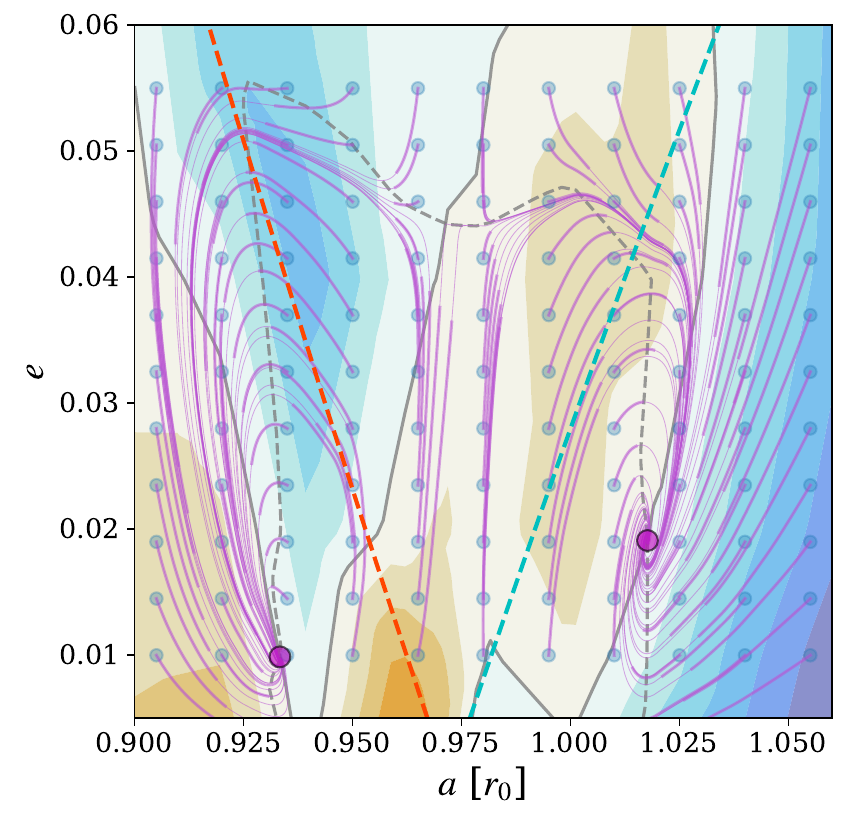}
  \caption{Same as Fig.~\ref{fig:traj}, except for a reduction factor
    of the luminosity of $25$\% (left) and $10$\% (right). Note that
    the axes limits are different on these two plots.}
  \label{fig:fid2510}
\end{figure*}

\begin{figure*}
  \includegraphics[width=\columnwidth]{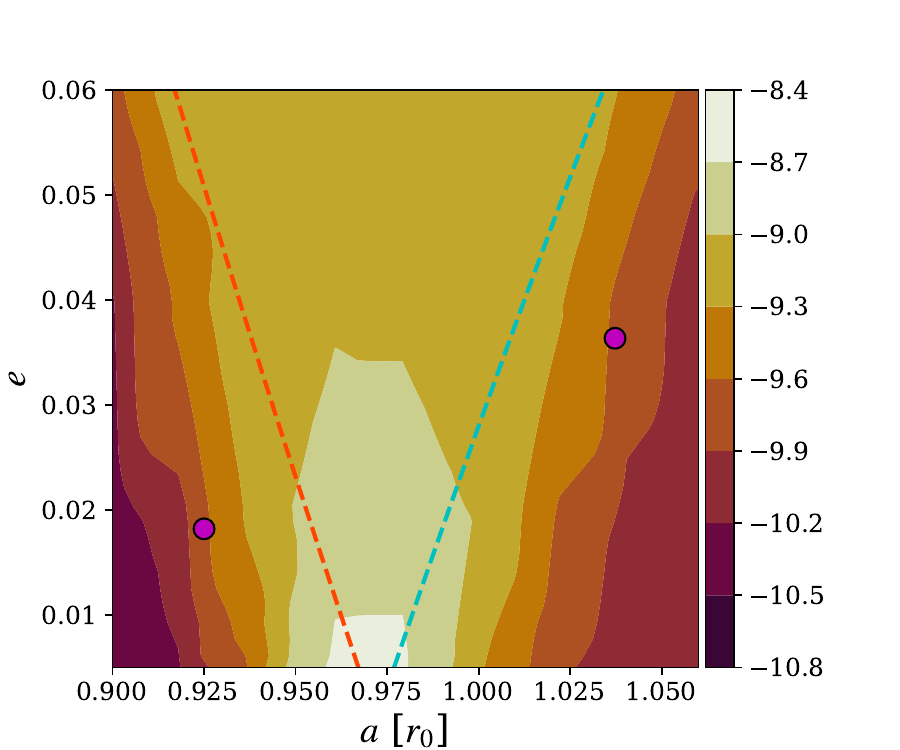}\includegraphics[width=\columnwidth]{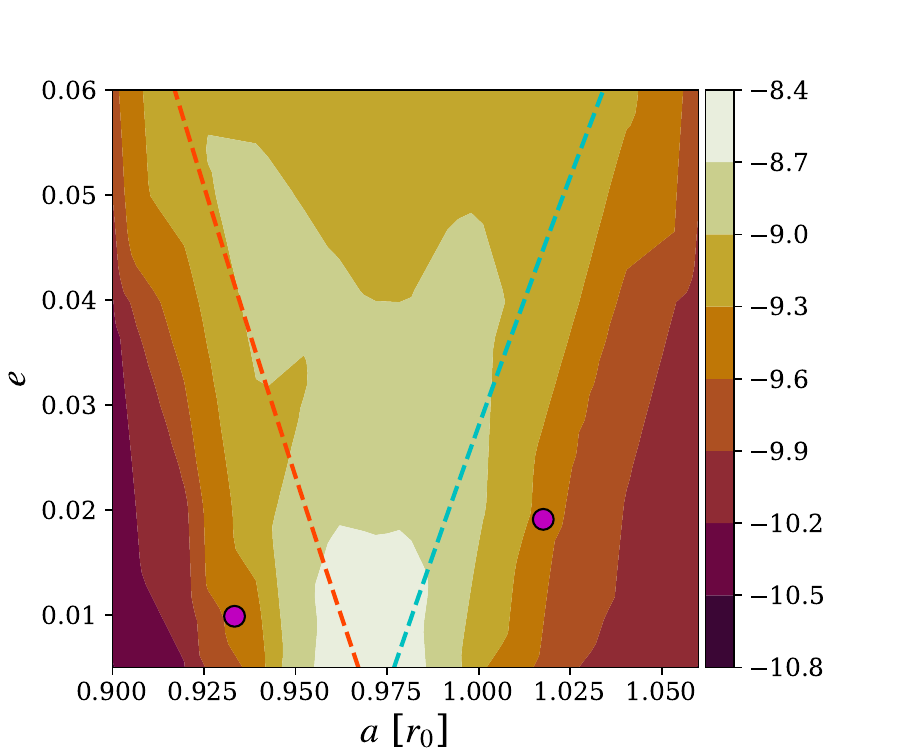}
  \caption{Same as , Fig.~\ref{fig:dmdtfidu} except for a reduction factor
    of the luminosity of $25$\% (left) and $10$\% (right). Note that
    the limits of the left plot have been adjusted to match those of
    the right plot.}
  \label{fig:dmdt2510}
\end{figure*}

We now deviate from our fiducial setup and repeat our calculations
with a different luminosity of the planetary cores. Namely, instead of
Eq.~\eqref{eq:42}, the luminosity is now given by:
\begin{equation}
  \label{eq:46}
  L=\varepsilon\frac{GM\dot M}{R_p},
\end{equation}
where $\varepsilon \in [0,1]$ is a reduction factor that accounts for
the fact that the pebbles accreted do not impinge on the surface of
the core and are instead vaporised before reaching the surface,
contributing to the formation of a high metallicity envelope
\citep{2018A&A...611A..65B}. While the mass accretion rate keeps same
value as before, the energy released in this case is smaller than the
value given by Eq.~\eqref{eq:42}, which can be regarded as a maximum
value. We have considered $\varepsilon = 0.25$ and
$\varepsilon=0.1$. For the sake of brevity, we do not reproduce here
the corresponding maps of $\dot e$ and $\dot a$, and directly show the
trajectories in the $(a,e)$ plane in Fig.~\ref{fig:fid2510} and the
map of mass accretion rate in Fig.~\ref{fig:dmdt2510}. We find that
the two fixed points identified in our fiducial exploration subsist,
with similar characteristics (one outside the ring and one inside, the
outside one having a larger eccentricity). As the reduction factor
$\varepsilon$ of the luminosity decreases, these points move toward
the ring and their eccentricity decreases. We see in
Fig.~\ref{fig:dmdt2510} that for a smaller reduction factor, the fixed
points have a location with higher accretion rate. Namely, we find
that for a $25$\% reduction factor, both fixed points correspond to a
very similar accretion rate with mass doubling time
$M/\dot M=13.6\cdot 10^4\Omega_p^{-1}$ or $69$~kyr for a central
object with a solar mass, significantly smaller than the mass doubling
time of the fiducial calculation. This trend to a smaller mass
doubling time continues for a $10$\% reduction factor, with a doubling
time for a planet at the inner fixed point of $10^4\Omega_p^{-1}$
orbits or $50$~kyr, and $8.8\cdot 10^4\Omega_p^{-1}$ or $44$~kyr at
the outer fixed point.

Examination of Figs.~\ref{fig:dmdtfidu} and~\ref{fig:dmdt2510} shows
that the mass accretion rate, for a given value of $a$ and $e$,
depends on $\varepsilon$: the smaller this value, the larger the
accretion rate. The radiative feedback from the planet onto its
immediate vicinity tends to evacuate the gas to larger distances. By
doing so, it also lowers the dust content that the planet can accrete.

\subsection{Effect at other planetary masses}
\label{sec:effect-at-other}
\begin{figure}
  \includegraphics[width=\columnwidth]{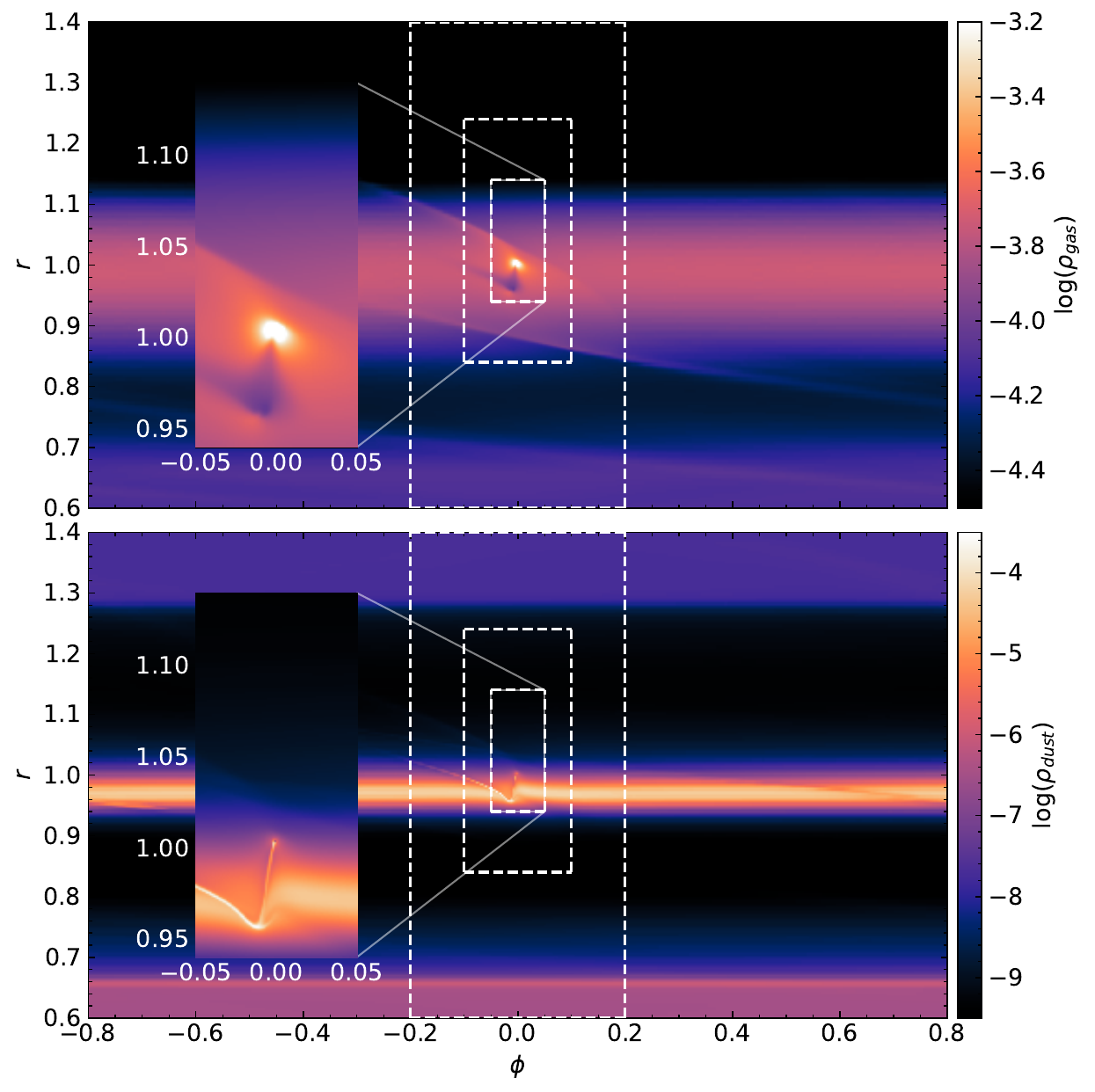}
  \caption{\label{fig:snapshot}Midplane gas density (top) and dust density (bottom) after the
    passage at periastron of a $4$~M$_\oplus$ planet with eccentricity
    $0.064$, semi-major axis $1.04r_0$ and accretion efficiency
    $\varepsilon=0.25$. The planet's orbital phase is $0.15$ (it would
    be $0$ at periastron and $0.5$ at apoastron. It is en route to
    larger radii and leaves in the gas a hot, under dense trail,
    clearly visible on the close up of the top plot. The planet in
    this setup has an orbit close to the fixed point (second larger
    green dot in Fig.~\ref{fig:fixedr}). Its radial excursion is
    marginally smaller than the extent of the layer of highest
    resolution, from $0.973r_0$ to $1.107r_0$. There is a hint of
    perturbations in the dust away from the planet, at azimuth
    $|\phi|\sim 0.6-0.7$.}
\end{figure}

\begin{figure}
  \centering
  \includegraphics[width=\columnwidth]{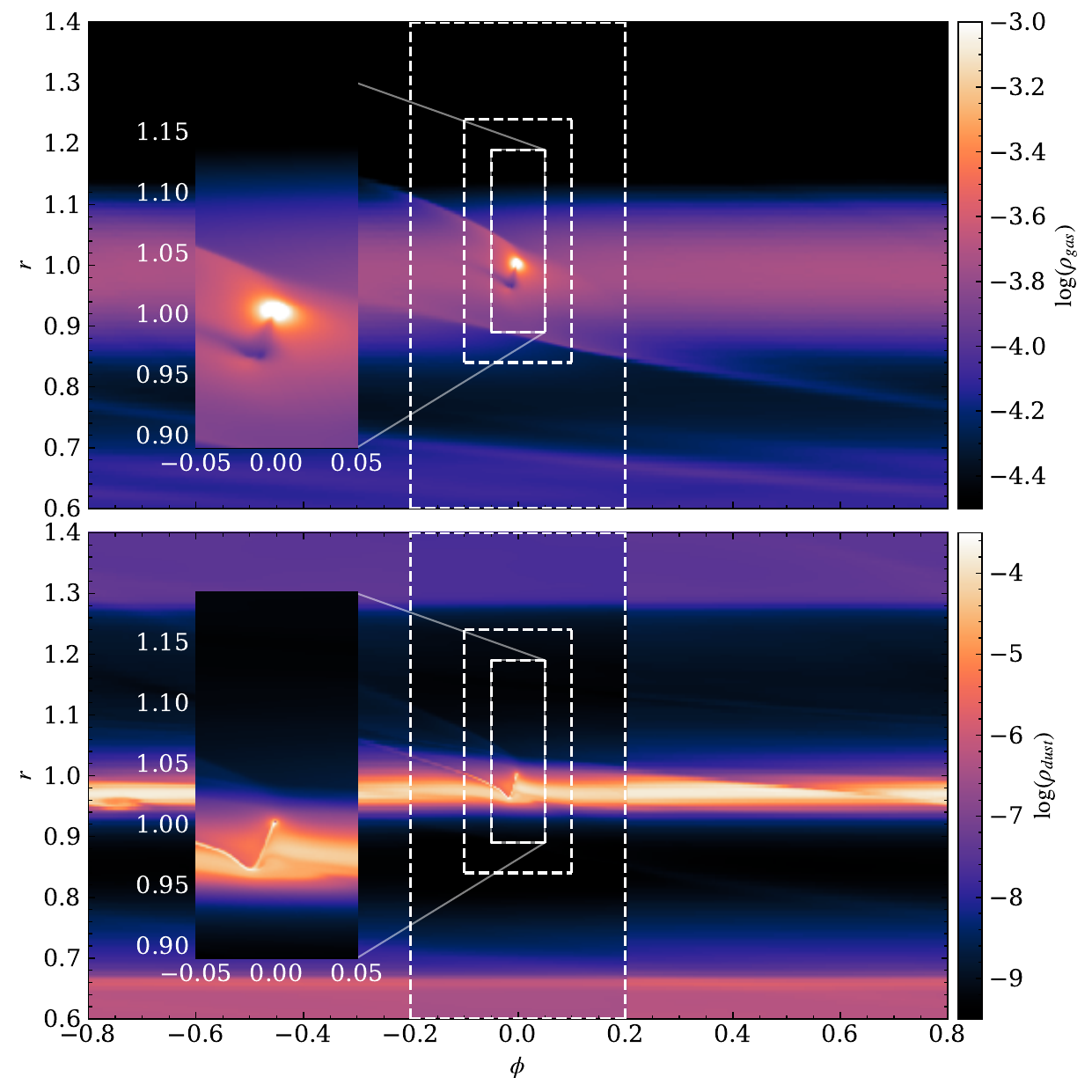}
  \includegraphics[width=\columnwidth]{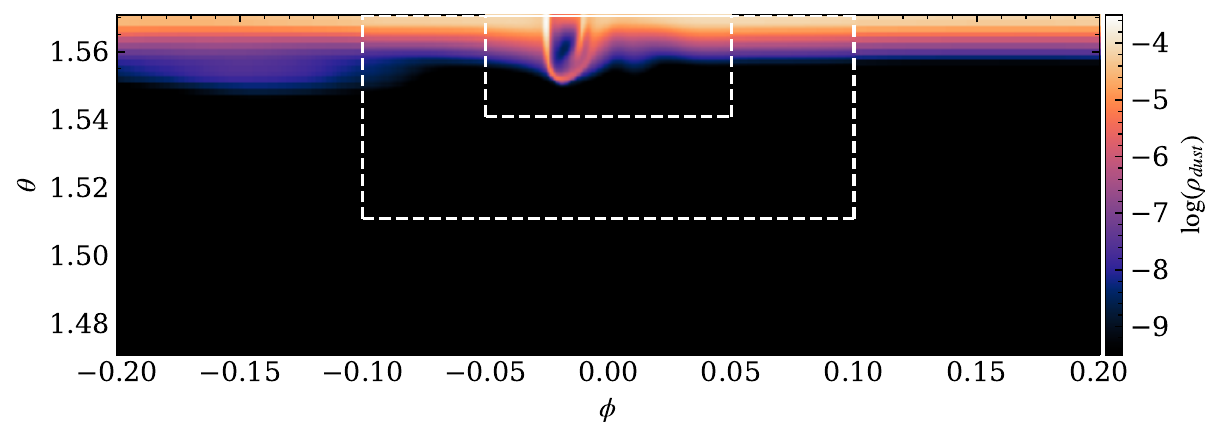}
  \caption{\label{snapshot2}Same as Fig.~\ref{fig:snapshot}, but for
    an $8$~M$_\oplus$ planet with eccentricity $0.06$, at orbital
    phase $0.14$. The other parameters are the same. The perturbation
    in the dust ring away from the planet is more apparent. It is due
    to a previous incursion of the planet. The bottom plot shows a cut
    of the dust density at $r=r_\mathrm{ring}=0.97r_0$. We see that
    the dust response is complex and markedly different from that of
    the gas. It shows an accumulation on a sheet in the region of low
    gas density, and a lift off the midplane.}
\end{figure}

We have investigated how the mechanism operates for planetary masses
different from the fiducial mass. Fig.~\ref{fig:snapshot} shows an
example of the gas and dust response at the disc midplane soon after
the passage at periastron of a $4$~M$_\oplus$ planet, while
Fig.~\ref{snapshot2} shows the response after the passage at periastron
of an $8$~M$_\oplus$ planet, both within the midplane and
vertically. We identified the fixed points in a number of cases.  In
order to limit the computational cost, we only explore small patches
of the $(a,e)$ plane where we expect to find the fixed points (and we
extend these patches in case it is necessary). We then measure the
mass accretion rate at each fixed point
identified. Tab.~\ref{tab:runsLR} summarises the different numerical
explorations that we have undertaken.

\begin{table}
\caption{Summary of the numerical explorations presented in this
  work. A letter W indicates that a wide exploration of the $(a,e)$
  parameter space has been performed (as presented in
  section~\ref{sec:fiducial-calculation}). A letter L (R) stands for a set
  of calculations that surround the left (right) fixed point. The
  case W$_f$ corresponds to the fiducial exploration.}
\label{tab:runsLR}
\centering
\begin{tabular}{cccccc}
\hline
\diagbox{$\varepsilon$}{$M$} & $0.1\;M_\oplus$& $1\;M_\oplus$&
                                                               $2\;M_\oplus$& $4\;M_\oplus$& $8\;M_\oplus$\\
\hline
  $0.1$ & $-$ & W & LR & LR & R \\
  $0.25$ & LR & W & LR & LR & R \\
  $1$ & LR & W$_f$ & $-$ & R & R \\
\hline
\end{tabular}
\end{table}
\begin{figure}
  \includegraphics[width=\columnwidth]{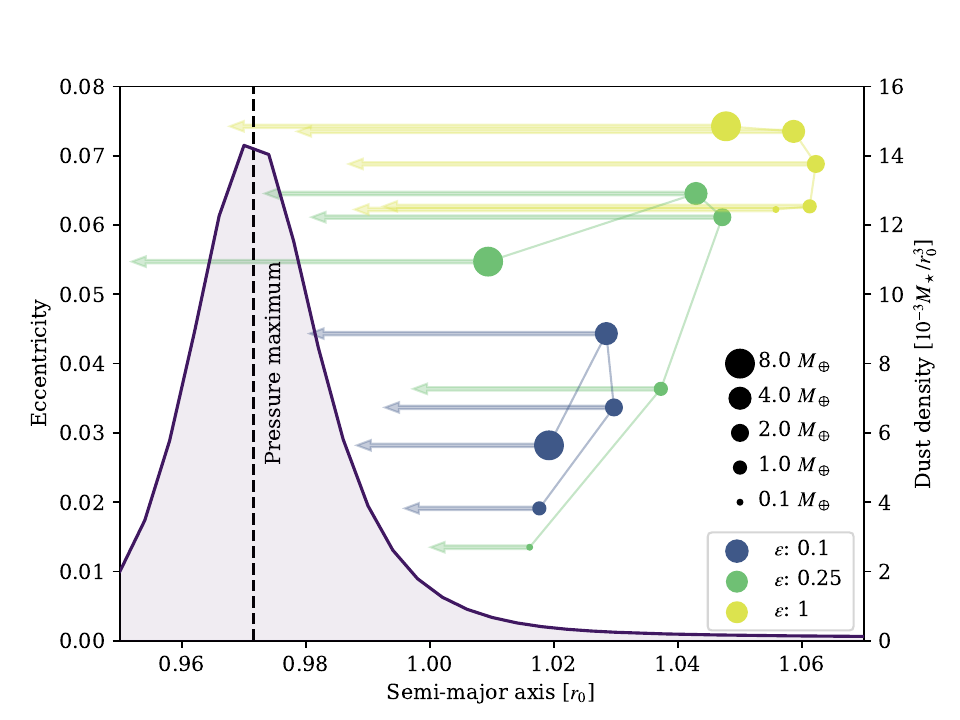}
  \caption{Location of the right fixed points.  The curve shows the
    midplane dust density. The arrows to the left indicate the
    location of the periastron $a(1-e)$. Most periastron distances
    fall where the dust density is a sizeable fraction of its peak
    value.}
  \label{fig:fixedr}
\end{figure}
\begin{figure}
  \includegraphics[width=\columnwidth]{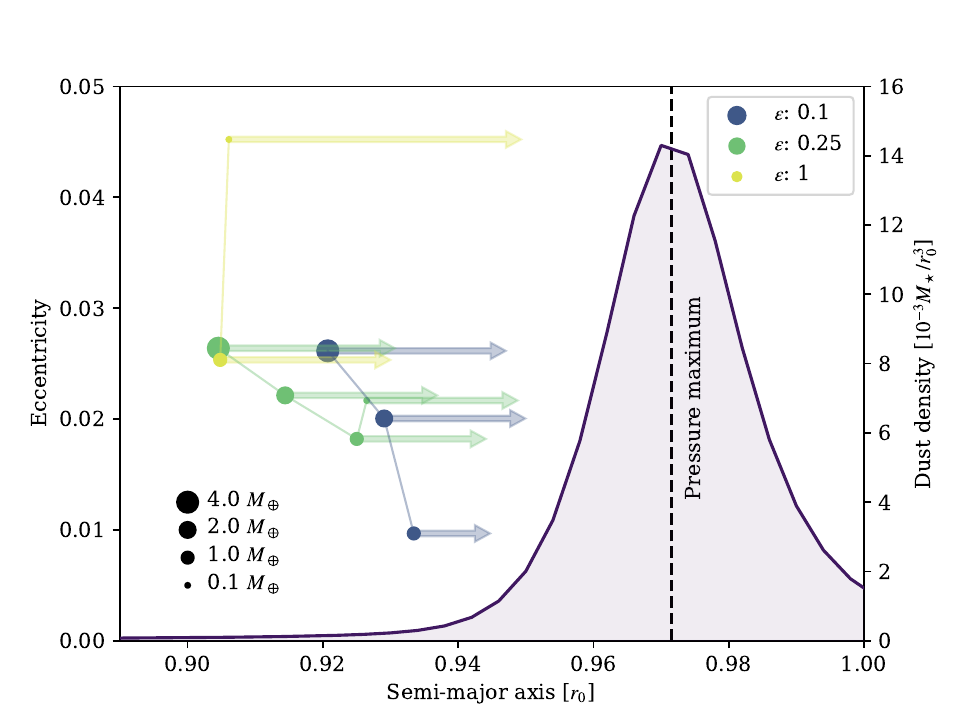}
  \caption{Same as Fig.~\ref{fig:fixedr} for the left fixed
    points. The arrows indicate this time the location of the
    apoastron $a(1+e)$. Contrary to Fig.~\ref{fig:fixedr}, the
    apoastron locations stay away from the peak, and at the closest
    distance of approach (that of the $2\;M_\oplus$ planet with
    $\varepsilon=0.1$), the density is $\sim 20$\% only of the peak
    value.}
  \label{fig:fixedl}
\end{figure}
We show in Fig.~\ref{fig:fixedr} and~\ref{fig:fixedl} the positions of
the right  and left fixed points, respectively.

The outer fixed
points, independently of $\varepsilon$, show a trend to move to
larger semi-major axis and higher eccentricity as the planet's mass
increases. For a planetary mass of order $\sim 4\;M_\oplus$, this
trend reverses and the semi-major axis decreases, while the
eccentricity can also decrease. We also see that the periastron
distance tends to move toward the centre of the dust's ring as the
mass increases, although this trend is not systematic. In two cases,
the planets of largest mass ($8\;M_\oplus$) cross the ring
centre.

The inner fixed point show similar trends with opposite sign for the
variation of the semi-major axis, except for the case $\varepsilon=1$,
for which we only have two points. The results that we have seen in
section~\ref{sec:vari-from-fiduc} generalise to all masses: the inner
fixed point has smaller eccentricity than the outer fixed point, and
the distance of closest approach to the ring is also larger for the
inner point.

\begin{figure*}
  \centering
  \includegraphics[width=.8\textwidth]{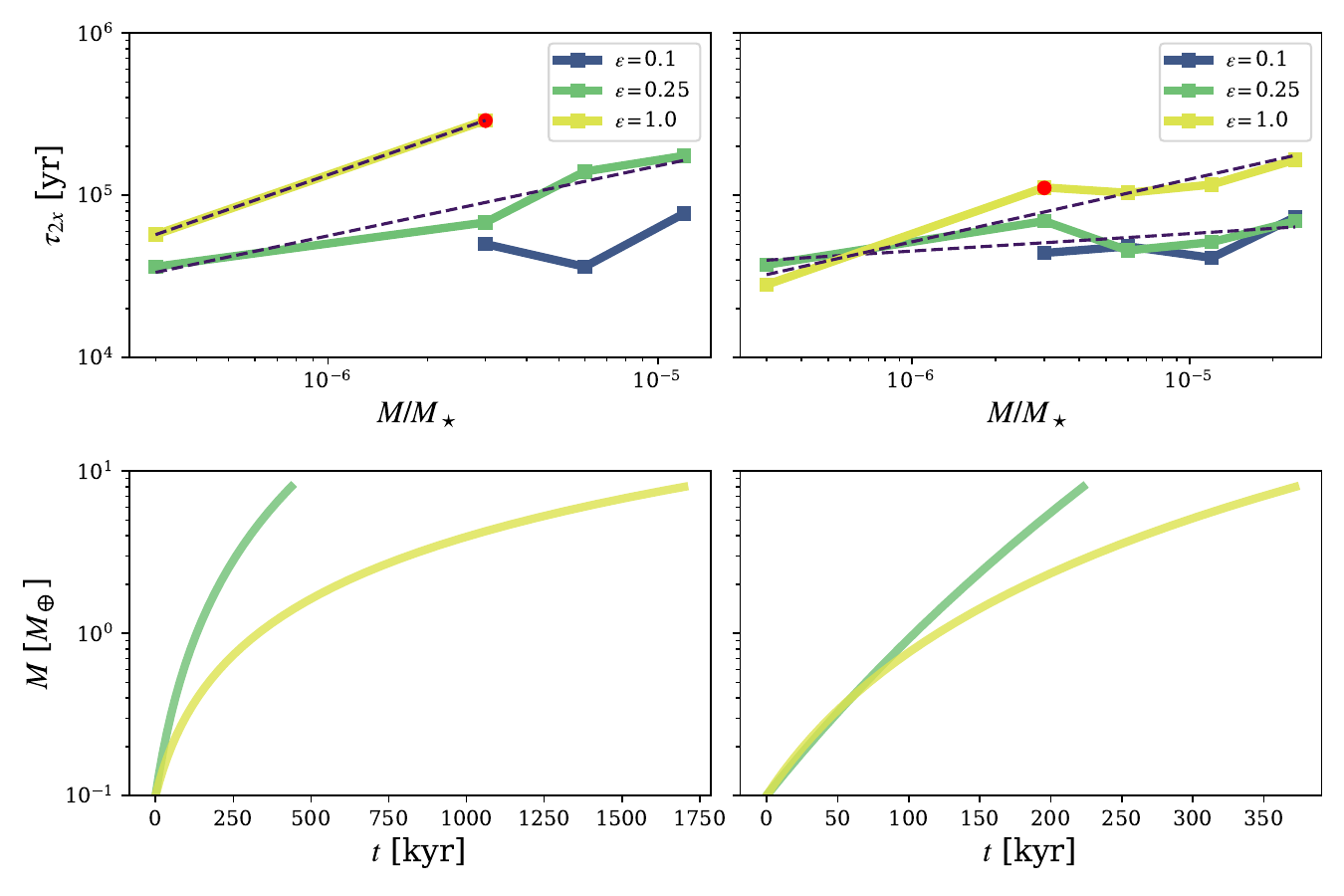}
  \caption{Top row: mass doubling time as a function of planetary mass
    at the inner (left) and outer
    (right) fixed points. Bottom row: planetary mass as a function of
    time, for a planet trapped at the inner fixed point (left) or at
    the outer fixed point (right). These time dependencies are
    analytical estimates obtained from the fits of the mass
    doubling time, shown with dashed lines in the first row. We do not
  provide a dependence for the case $\varepsilon=0.1$, for which we do
not have the lowest mass $M=0.1\;M_\oplus$. 
}
  \label{fig:mvst}
\end{figure*}

Fig.~\ref{fig:mvst} shows the mass doubling times obtained for the different fixed points studied in the present work. A fit of these times is subsequently used to provide the planetary mass as a function of time for different values of $\varepsilon$, at the inner and outer fixed points, up to the largest mass considered in our study ($8\;M_\oplus$), which we can consider representative of the critical mass for runaway gas accretion. Since the mass doubling times have been obtained for a ring with $10$~$M_\oplus$ of dust, the time evolution of the planetary mass is representative of the case in which the ring is permanently replenished by inwardly drifting dust so as to keep its mass at $10$~$M_\oplus$. We recover the fact that the mass grows faster for a smaller value of $\varepsilon$, and that a trapping at the outer orbit favours a faster growth. For the least favourable case, that of a planet trapped on an  inner orbit with $\varepsilon=1$, it takes almost $2$~Myr for the planet to reach $8\;M_\oplus$, while in the most favourable case, that of a planet trapped on an outer orbit with $\varepsilon=0.25$, this growth is completed in only $200$~kyr.

\section{Discussion}
\label{sec:discussion}
\subsection{Growth inside or outside the ring: a possible dichotomy}
\label{sec:growth-inside-or}
\begin{figure}
  \centering
  \includegraphics[width=\columnwidth]{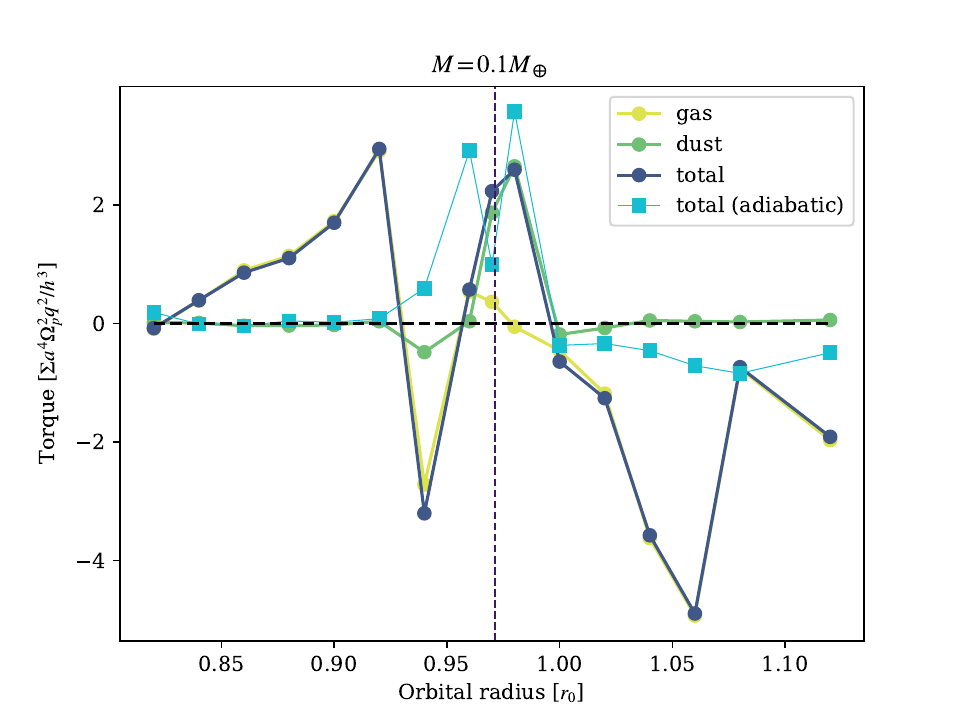}
  \caption{Normalised torque exerted on a non-luminous low-mass embryo
  near the ring. The gas dominates the total torque, except within the
ring where the dust nearly accounts for all the torque. The thin line
with squares shows the total torque when the gas is adiabatic.}
  \label{fig:tqcircmars}
\end{figure}

We have found that protoplanets in the vicinity of a dusty ring can be
trapped either on an inner eccentric orbit or an outer eccentric
orbit. We have also found that in general, an outer eccentric orbit
goes deeper into the ring (at periastron) than does an inner eccentric
orbit (at apoastron), with the consequence that the mass growth is
significantly slower for a planet trapped on an inner eccentric
orbit. The ultimate fate of a very low mass core forming in the ring
therefore depends strongly on which of the two fixed points in $(a,e)$
space it will eventually reach. If it goes to the outer fixed point,
it may reach the critical mass for runaway gas accretion in a few
$10^2$~kyr (for our disc parameters and a ring at $10$~au), whereas
it may remain a super-Earth if it goes to the inner fixed point. We
have performed additional calculations of a non-luminous Mars-sized
embryo on a circular orbit at different orbital radii in the ring. We
have seen in the previous sections that even a Mars-sized embryo would
have its eccentricity excited by the feedback of its accretional
luminosity. Setting here the luminosity to zero gives therefore an
indication of how the torque depends on the orbital radius for an even
lower mass planet, with a subcritical luminosity, which is still on a
circular orbit. We see in Fig.~\ref{fig:tqcircmars} that the total
torque cancels out at several radial locations. When the radial
derivative of the torque is negative, such a location constitutes a
trap (whereas if the derivative is positive, the location is unstable,
as can be easily checked). The two trap locations are found here near
$r=0.93r_0$ and $r=r_0$, and the torque at the pressure maximum is
positive, which suggests that an embryo born at the middle of the ring
will eventually get trapped near $r_0$. As the planet mass grows, it
becomes luminous. Initially it has a subcritical luminosity ($L<L_c$)
and thus remains on a circular orbit. As the luminosity grows, the gas
torque is no longer dominated by the cold thermal torque
\citep{2014MNRAS.440..683L}, and at the point at which the luminosity
becomes critical and eccentricity starts growing, the heating and cold
thermal forces cancel each other, so that the torque has nearly its
adiabatic value \citep{2017MNRAS.472.4204M}. We also plot in
Fig.~\ref{fig:tqcircmars} the total torque when the gas is
adiabatic. It still shows a trap near $r_0$, which suggests that as
the embryo will start growing its eccentricity, it will eventually
reach the outer fixed point. Nonetheless, owing to the complexity of
the torque behaviour with radius, it sounds plausible that, depending
on the circumstances, a planet could also reach the inner
point. \citet{2024arXiv240205760P} find that a planet growing near the
dust peak eventually migrates inwards or outwards, depending on
whether they include the feedback of the dust onto the gas (and they
find, with feedback, the result that we find here without
feedback). This underlines the extreme sensitivity of the path of
the planet to the details of the physical processes responsible for
the torque, and suggests that a bifurcation can occur when the planet
starts becoming eccentric, with considerable consequences on its
destiny. Assessing which of the two fixed points a growing embryo will
eventually reach in a general case warrants significant further work.
 
\subsection{Frequency mismatch between the planet and the ring:
  consequences}
\label{sec:freq-mism-betw}
Every time the eccentric planet makes an incursion in the ring, it
accretes dust and leaves a ``scar'' on the ring, with a typical width
equal to the radius $r_\mathrm{acc}$ of pebble accretion. At the next
passage, owing to the frequency mismatch between its orbital frequency
and the orbital frequency in the ring at the radius of the planet's
incursion, the planet accretes from another, unperturbed region of the
ring. The time it takes for the first perturbation to coincide again
with the planet is equal to the synodic period of the ring seen by the
planet, and is of order of:

\begin{equation}
  \label{eq:47}
  T_\mathrm{syn}\sim\frac{2\pi}{|r\partial_r\Omega|e}=\frac{4\pi}{3e}\Omega
  ^{-1}
\end{equation}
This value is to be compared with the time it takes to erase the
scar by dust diffusion. We make hereafter a conservative derivation of
this time, in which we neglect the Keplerian shear.
\begin{equation}
  \label{eq:48}
  T_\mathrm{diff}=\frac{r_\mathrm{acc}^2}{\delta c_s^2/\Omega}.
\end{equation}
The accretion radius has to be evaluated in the headwind regime,
appropriate for an eccentric planet
\citep[e.g.][]{2017ASSL..445..197O}:
\begin{equation}
  \label{eq:49}
  r_\mathrm{acc}=\sqrt\frac{2GM\tau_s}{(1/2)er\Omega^2}=2r\sqrt\frac{q\tau_s}{e}.
\end{equation}
Using Eqs.~\eqref{eq:48} and~\eqref{eq:49}, we can recast the
diffusion time as:
\begin{equation}
  \label{eq:50}
  T_\mathrm{diff}=\frac{4q\tau_s}{\delta eh^2}\Omega^{-1}.
\end{equation}
By comparison with the synodic period, we see that the ``scar'' is
erased at the next passage of the planet if:
\begin{equation}
  \label{eq:51}
  \frac{q\tau_s}{\delta h^2}<\pi\mbox{~~or~~}\frac{q}{h_d^2}<\pi.
\end{equation}
This condition simply consists of a comparison between the
planet-to-star mass ratio and the square of the aspect ratio of the
dusty disc. With our parameters, it translates into:
\begin{equation}
  \label{eq:52}
  q \lesssim 7\times 10^{-5}\mbox{~~or~~}M\lesssim 26\;\mathrm{M}_\oplus.
\end{equation}
This mass limit is above the threshold for runaway gas accretion.  The
vertical cut of dust density for an $8$~$M_\oplus$ planet displayed in
Fig.~\ref{snapshot2} shows that the dust is lifted off the
midplane. The time it takes for the dust particles to settle back
toward the midplane is $\sim(\Omega_K\tau_s)^{-1}$, which for our choice
of the Stokes number amounts to $16$~orbital periods, comparable to
the synodic period for a planet with an eccentricity comparable to the
gaseous disc's aspect ratio.

It is therefore reasonable to consider that, in our ring, a rocky
protoplanetary core essentially accretes from an unperturbed dusty
ring at each incursion. If the diffusion parameter $\delta$ is
smaller, or if the dimensionless stopping time of the pebbles is
larger, the dusty disc may be much thinner and the mass limit may be
significantly smaller. It would then be necessary to take into account
the fact that the planet accretes from a dusty disc significantly
perturbed by its previous passages. Figs.~\ref{fig:snapshot}
and~\ref{snapshot2} reveal mild perturbations of the ring away from
the planet, especially in the case of an $8$~M$_\oplus$ planet. In
these simulations, however, the setup has an azimuthal period of
$1.6$~rad, which entails a shorter synodic period and less time for
the ring to relax toward the unperturbed configuration between two
successive incursions of the planet.

The diffusion time $\tau_\mathrm{ring}$ of the dust across the ring is
relatively short. Using Eq.~\eqref{eq:37} and the fact that
$w_g\gtrsim H$, we obtain
$\tau_\mathrm{ring} \sim w_d^2/(\delta H^2\Omega_\mathrm{K}) \gtrsim
(\tau_s\Omega_\mathrm{K})^{-1}$, comparable to the vertical settling
time. This timescale being much shorter than that of accretion, if the
ring is not fed from outside by an inward flux of pebbles, the surface
density of dust decays uniformly while keeping a profile similar to
the initial one. The fixed points move to adjust to the slowly varying
surface density of the ring, and, as long as they exist, the planet
keep feeding from the ring. We remark that the impact of the decrease
of the ring's surface density on the position of the fixed points
should be exactly the same as that of the decrease of $\varepsilon$
presented in section~\ref{sec:vari-from-fiduc}: the fixed points
should move toward the ring.  The trapping on an eccentric orbit
ceases when the dust density is low enough that the planet's
luminosity becomes subcritical and its eccentricity is damped. The
whole process should be studied (i) either via long-term numerical
simulations at intermediate resolution (the resolution used in the
present study does not allow simulations over thousands of orbits over
a reasonable timescale) possibly with a different prescription for
dust accretion than the one used here, as coarser cells near the
planet may be larger than its effective accretion radius; (ii) or in a
semi-analytical manner, by tracking the evolution of the azimuthally
averaged dust density in the ring.

We comment that the synodic period of the planet with respect to the
location of the ring where the accretion rate peaks is relatively
short ($\gtrsim 10$~orbits) compared to the timescale of the
variations of semi-major axis and eccentricity ($\sim
10^2$~orbits). Consequently, if there are azimuthal variations of the
dust surface density along the ring (hence variations of the peak
luminosity and thermal force from one incursion to the next), their
impact on the orbital evolution of the planet should average out and
the trapping process should be broadly similar to that for an
axisymmetric dust ring. We therefore expect a dust ring subjected to
the RWI \citep{lovelace99,li2000} or the dusty RWI
\citep{2023MNRAS.526...80L} to trap rocky planets on eccentric
orbits. For the same reason we expect that the trapping mechanism
would essentially be unchanged should the ring be mildly perturbed
from previous incursions of the planet.

\subsection{Ring expansion}
\label{sec:ring-expansion}
The process presented here relies almost exclusively on thermal
torques, as discussed in length in
section~\ref{sec:torq-revers-diff}. The usual resonant torques
(Lindblad and corotation) are immaterial for the trapping. All what
matters is that the planet can feed from a radially narrow
distribution of dust, in order to have a significantly variable
luminosity. For the needs of the present study, we realised this
radially narrow dust ring by trapping dust within a pressure bump, but
\emph{per se} the pressure bump is not required for the trapping at the outer orbit to
take place, as mentioned in section~\ref{sec:interpr-results}. It should also occur in the clumpy rings proposed by
\citet{2021MNRAS.505.1162J}, which have sharp edges (at least on their
outer edge). A remarkable property of these rings is their trend to
expand with time. Could a planet trapped on an eccentric orbit by such
a ring follow its expansion, or would it be left in place ? A planet
that would follow the ring's expansion would no longer have
$(\dot a,\dot e)=(0,0)$. In the frame comoving with the ring, a steady
state would be achieved for
$(\dot a, \dot e)=(\dot r_\mathrm{ring},0)$. The location of the
planet on the $\dot a$ map would then be given by the intersection of
the $\dot e=0$ contour and that with $\dot a=\dot r_\mathrm{ring}$, if
it exists. For an Earth-mass object trapped on an outer orbit, we read
in Fig.~\ref{fig:deda} that the maximal achievable drift rate on the
$\dot e=0$ contour would be of order of $10^{-4}
(GM_\star/a)^{1/2}$. This corresponds to
$\dot a_\mathrm{max} \approx 2\times
10^{-4}$~au.yr$^{-1}=200$~au.Myr$^{-1}$.  This maximal rate should be
compared to the expansion rates found by
\citet{2021MNRAS.505.1162J}. They find a particularly fast expansion
for a ring formed at $10$~au (their Fig.~12), with an initial rate
$\lesssim 50$~au.Myr$^{-1}$, while their Tab.~1 reports drift
velocities that are all below this maximal rate (in general by more
than one order of magnitude, and marginally for one of
them). Naturally, the characteristics of these rings differ from those
of the ring considered in this study, but given the considerable
margin we find even for our lightweight ring ($10$~M$_\oplus$), the
mechanism we report here should allow expanding clumpy rings to easily
transport growing planets to several tens of astronomical unit over
timescales $\gtrsim 1$~Myr, provided the mechanism unveiled here
remains efficient at larger orbital distances.

This discussion begs the question of how far from the star can
  the trapping mechanism operate. It essentially boils down to a
  comparison of the planet's luminosity achievable at a given distance
  to the local value of the critical luminosity, given by
  Eq.~\eqref{eq:2}. When the former is much larger than the latter,
  the thermal forces at the passage during the ring dominate the
  dynamical evolution of the planet. For a planet of given mass, given
  eccentricity and for a given Stokes number, the accretion radius
  scales with the orbital distance of the planet $r$, while the
  planet-dust relative speed scales with $r^{-1/2}$. The luminosity
  achievable therefore scales with $r^{1/2-\alpha_r}$ regardless of
  whether accretion is in the 2D or 3D regime, where $\alpha_r$ is a
  slope of surface density similar to that introduced in
  section~\ref{sec:initial-conditions}, describing how the dust
  surface density in the rings vary with radius. The critical
  luminosity itself scales with two key quantities, that have opposite
  behaviour with radius: the thermal diffusivity, which increases
  outwards, and the density of the gas at the midplane (which
  decreases outwards). The thermal diffusivity depends itself on the
  temperature, density and opacity as described by
  Eq.~\eqref{eq:29}. Assuming the latter scales as $\rho^0T^2$
  \citep{1994ApJ...427..987B}, the critical luminosity scales as
  $r^{3f+\alpha}$, where $f$ is the disc's flaring index
  $\partial \log h/\partial\log r$. The ratio of the planet's
  achievable luminosity to the critical one therefore scales as
  $r^{1/2-\alpha_r-\alpha-3f}$. Unless the decay of surface density in
  the rings is very shallow and the disc has little flaring, this
  quantity decreases with radius and the mechanism presented should
  cease to operate at some distance from the central object. We have
  seen that by reducing the luminosity by a factor of ten, the
  trapping is still active. If we therefore take a factor of ten as a
  conservative estimate, and scale the properties of the ring
  considered here, we infer that the trapping would become
  inefficient, for a nominally luminous planet with $\varepsilon =1$,
  at a radius
  $r_c\sim 10\mbox{[au]}\times 10^{1/(3f+\alpha_r+\alpha-1/2)}$. As an
  example, if $\alpha=\alpha_r=1/2$ and $f=0.25$, this expression
  yields $r_c\sim 63$~au., while if $\alpha=\alpha_r=1$ and $f=0.25$,
  it yields $r_c\sim 28$~au. Note, however, that the dust \textit{of a
    given size} has a larger Stokes number at larger distance from the
  star \citep{2014PhDT.......329D}. We noticed in section~\ref{sec:governing-equations} that
  larger Stokes numbers are more favourable to the trapping
  mechanism. This could allow the mechanism to operate at distances
  significantly larger than the limits quoted above.

\subsection{Comparison to previous work}
\label{sec:comp-prev-work}
In recent years, there has been a substantial amount of work on the
formation of planets in dust rings. \citet{2020A&A...638A...1M}
considers the growth of initially Mars-sized embryos in the vicinity
of dusty rings at pressure bumps, both at large distance from the
central object ($75$~au) and at small distance
($5$~au). \citet{2020A&A...642A.140G} consider the growth of
planetesimals and planets in a pressure bump at the ice line
($3$~au). \citet{2021ApJ...914..102C} considers the formation of
planets in several, fixed pressure bumps with radii in geometric
sequence, starting from pebble accreting planetesimals, all the way to
gas accreting, large mass planets. \citet{2022A&A...668A.170L} study
the formation of rocky cores at pressure bumps, both relatively close
to ($\sim 14$~au) and far from ($\sim 100$~au) the central object,
starting from dust coagulation and drift. \citet{2023MNRAS.518.3877J}
study planet formation in dusty rings, starting from the formation of
planetesimals, which subsequently accrete pebbles. These authors do not
limit themselves to dust rings at pressure bumps, unlike previous
studies, and also incorporate results about the formation of planets
in clumpy rings \citep{2021MNRAS.505.1162J}, which do not rely on the
presence of a pressure maximum. Broadly, these results show that the
assembly of rocky cores with a mass sufficient to trigger runaway gas
accretion occurs on timescales shorter than lifespan of the disc in
the inner disc ($r \lesssim 10$~au), while the results are more
nuanced in the outer disc ($r =
O(10^2)$~au). \citet{2020A&A...638A...1M} finds unlikely the build up
of critical mass cores at $75$~au, essentially because the cores carve
a gap in the dust and have their accretion rate limited by the dust
diffusion. \citet{2021ApJ...914..102C} subsequently argues that the
moderate eccentricity of the cores, in addition to their finite
accretion radius, allow them to explore a more extended region,
thereby increasing their accretion rates. Some of their calculations
show indeed giant planet at large
distances. \citet{2022A&A...668A.170L} obtain critical mass cores on
short timescales (of the order of 100~kyr) even at large distances
from the star, while \citet{2023MNRAS.518.3877J} similarly obtain
$10$~$M_\oplus$ cores in clumpy rings at $75$~au, on timescales
shorter than $1$~Myr. Crucial to the outcome is the torque exerted on
the planet during the phase from $1$ to $10$~$M_\oplus$.  In this
respect, these studies share many similarities. They only consider the
torque exerted by the gas and discard that from the dust. The latter,
however, has been shown to be potentially important even in smooth
discs \citep{2018ApJ...855L..28B,2023ApJ...953...97G}, and is dominant
in pressure bumps \citep[][and the present work ---~section
\ref{sec:growth-inside-or}]{2024arXiv240205760P}. The torque from the
gas is computed from an \textit{ad hoc} density profile for the
pressure bump, which is in general Gaussian. Several degrees of
accuracy are considered. Some authors use the torque formulae for
locally isothermal discs obtained for discs with power law profiles of
surface density and temperature, and inject in these formulae the
rapidly varying slopes of surface density and temperature within the
bump to obtain torque expressions. Others consider non-isothermal
effects and the saturation of the corotation torque, for which they
use more sophisticated torque formulae
\citep{pbk11,2017MNRAS.471.4917J}, or a variant
\citep{2021ApJ...914..102C} in which non-isothermal effects are
accounted for \citep{pbck10}, but not the saturation of the corotation
torque. The migration path of the planets is then bracketed between
that obtained with such prescription, and one obtained from the
Lindblad torque only, corresponding to a fully saturated corotation
torque. Taking into account the saturation of the corotation torque
(i.e., its trend to decay towards a small value) is of particular
importance: the Lindblad torque alone cannot halt migration, as it is
invariably negative \citep{2010ApJ...724..730D}, even at pressure
bumps \citep{2011CeMDA.111..131M}. The corotation torque is therefore
required to obtain a planet trap, but this can only happen if it is
not saturated. Unless the turbulence is very weak (i.e. the effective
viscosity is very low), a significant corotation torque should subsist
for protoplanets with masses up to that of critical mass cores
($\sim 10$~M$_\oplus$), but the degree of saturation of the torque
regulates the location of the trap with respect to the peak of dust
and ultimately the accretion rate of pebbles
\citep{2020A&A...638A...1M}. In addition to the Lindblad and
corotation torque, the planet is subjected to thermal
torques. \citet{2020A&A...642A.140G} include these torques, but assume
the planet to be on a circular orbit independently of whether its
luminosity is sub- or super-critical, and they abruptly set thermal
torques to zero passed the relatively small critical mass
\begin{equation}
  \label{eq:53}
  M_c=\frac{\chi c_s}{G}.
\end{equation}
As a consequence of their assumption of a circular orbit, they cannot
observe the effects we report here.  An additional effect not taken
into account in these studies is the role played by dynamical
corotation torques \citep{2014MNRAS.444.2031P,2015MNRAS.454.2003P} on
the rocky cores, or their higher mass version \citep{mp03}. As there
seems to be a shift of paradigm toward laminar discs dominated by
magnetised winds \citep{Bai_2013,2017ApJ...836...46B}, in which
accretion is driven by magnetic torques, dynamical corotation torques
can build up even on slowly or non-migrating planets
\citep{2018MNRAS.477.4596M} in the Earth mass range and completely
alter their orbital evolution.

Notwithstanding the fact that a growing planet in the
$O(10^{-1})-O(10)\;M_\oplus$ range should be on an
eccentric orbit rather than a circular one, it can be seen that the
determination of the accretion rate of a planet on a circular orbit in
a dusty ring is a very complex problem, which requires a detailed
knowledge of the ring profiles and of the microphysics in the ring to
establish accurately the distance of the orbit to the peak of dust. This
distance may be, or not, favourable to the planet growth, and a given
scenario of growth and migration is inherently uncertain anyway due to
the neglect of dynamical corotation torques.

The mechanism we present here is completely immune to these issues. It
occurs whenever the dust density has relatively narrow radial
variations, it is virtually independent of Lindblad's and corotation
torques, and allows for a systematic consumption of the dust until the
planet's luminosity becomes subcritical. The question of the location
of the orbit becomes that of the location of the periastron (or
apoastron), and that location is precisely dictated by the
concentration of dust, rather than by the subtle balance of resonant
torques from the gas.

Very recently, \citet{2023MNRAS.524.2705C} investigated the evolution
of low-mass planets near pressure bumps. Those with subcritical
luminosities, predicted to be trapped near the pressure local maximum
\citep{2017MNRAS.472.4204M}, are indeed found to remain at the
pressure bump. Once their luminosity becomes supercritical, they
become eccentric. However, their luminosity is kept constant, rather
than being modulated according to the underlying dust density. Owing
to the lack of variation of the heating force along their epicycle,
they suffer the fate described in section~\ref{sec:torq-revers-diff}:
they migrate inwards and escape the ring. \citet{2024arXiv240205760P}
studied the growth and orbital evolution of low-mass planets at
pressure bump through extensive two-dimensional simulations, which
allow much longer integrations than our three-dimensional simulations
with nested meshes.  They release the accretional luminosity to the
gas in the vicinity of the planet, and have a realistic prescription
for the accretion of dust, modelled as a pressureless fluid. These
prescriptions allow for a variation of the luminosity (and heating
force) along the orbit. They observe behaviours similar to the ones we
report, which we believe to be based on the same mechanism. Namely,
they observe that the planet can settle outside the ring with an
eccentricity $e\sim h$ and a variable accretion rate, much larger at
periastron, compatible with a planet trapped at an outer fixed
point. They also find instances of an eccentric planet trapped inside
the ring, such as the case without dust back-reaction on their
Fig.~8. The overshoot of eccentricity, and the initial slow decay of
the semi-major axis, are typical of the trajectories about the
inner fixed point (see Fig.~\ref{fig:traj}), and suggest that the mechanism
at work in their simulations is similar in nature to the mechanism
reported here.  Thanks to the long-term nature of their calculations,
they also observe interesting additional effects that our short runs are
unable to capture, such as the formation of a dust-vortex and its
interaction with the planet, which warrants further work.

\subsection{Caveats of our analysis}
\label{sec:caveats-our-analysis}
We draw here a non-comprehensive list of the caveats of the present
analysis.
\subsubsection{Only one, non-inclined planet}
\label{sec:only-one-non}
In the present work, we have considered only one planet at a time,
assumed to be coplanar with the disc. If there is indeed only one
planet, the assumption of coplanarity is a reasonable one:
\citet{2017arXiv170401931E} have found that the eccentricity grows
$\sim 3\times$ times faster than inclination, and that once the
eccentricity reaches significant levels (larger than $\lambda/a$), the
growth of inclination stops, so that the inclination remains at a
very small value. However, if various embryos with super-critical
luminosity are simultaneously present on the same side of the ring
(either inside or outside), they may undergo close encounters which
will change their eccentricity and inclination. The evolution of an
embryo with an inclination larger than the aspect ratio of the dusty
disc depends on the argument of periastron: if the latter is close
to~0 or 180$^\circ$ (i.e. if it lies near the line of nodes), the
planet goes through the dust at periastron, and effects similar to
those described here should occur. The dynamics in this case warrants
further study as the introduction of a new degree of freedom, the
inclination, may lead to significant changes with respect to the
scenario of a coplanar planet. If the line of nodes and periapse are
misaligned, the planet does not accrete significantly on any part of
its orbit. Its luminosity drops and, if it becomes sub-critical, the
inclination and eccentricity are damped. The subcritical planet should then resume
a migration toward the ring \citep{2023MNRAS.524.2705C}, until it
accretes a sufficient amount of dust to repeat the whole process of
convergence toward the fixed point in $(a,e)$.

\subsubsection{No feed back of the dust}
\label{sec:no-feed-back}
In this exploratory work we have not considered the feed back of the
dust onto the gas in the present work. The dust to gas ratio
near the centre of the ring suggests that the dust feedback onto the
gas, which is neglected in the present analysis, may play a role. The
streaming instability could set in, and the formation of planetesimals
could occur in the ring. We do not take into account this process, nor
the accretion of planetesimals, in our analysis. \citet{2024arXiv240205760P}
find that embryos follow different paths depending on whether the feed
back is included. The inclusion of feed back should be important in
determining the dust torque while the planet has a small
eccentricity. When the planet has a sizeable eccentricity and is near
a fixed point, the feed back should not have a strong impact on the
net force, then dominated by the thermal force.  We do not address either
the question of the life expectancy of the ring, which is beyond the
scope of this paper. Should the lifetime of a ring be shorter than the
time it takes for the planet to accrete most of its mass, the trapping
mechanism envisioned here would stop, as the accretion rate would drop
below the levels required to counteract migration. If the residual
accretion rate drops below that corresponding to the critical
luminosity of the planet (Eq.~\ref{eq:2}), its eccentricity would
decay and its inwards migration would resume. However, given that most
discs do exhibit several rings at a given time
\citep{2018ApJ...869L..42H}, it could approach another ring from the
outside and end up trapped on the outer side, resuming accretion,
unless there is no ring inside to prevent its migration. This
hypothetical scenario underlines that planet's growth and orbital
evolution is tightly linked to the evolution of the rings, not only
because they feed from them, but also because their evolutionary path
strongly depends on the time behaviour of the rings' radii and their
life expectancy.

\subsubsection{No cut-off of the pebble accretion}
\label{sec:no-cut-pebble}
There is a variety of effects that may limit the ability of the planet
to accrete pebbles which have not been taken into account in our
approach. Our accretion procedure, which removes pebbles from the
8~zones nearest to the planet, does not consider the turbulent
stirring that may inhibit the settling of a fraction of the dust
content.

Planets with eccentricities larger than the disc's aspect ratio have a
supersonic motion with respect to the gas on the radial parts of their
epicycle, and have therefore a bow shock. Pebbles crossing the bow
shock may be destroyed \citep{2018A&A...615A.138L}. We do not take this
effect into account. Note however that in most cases the planet does
barely accrete except at peri- or apoastron. At those locations, its
velocity with respect to the gas is half the velocity it has when it
crosses the circle of radius equal to the semi-major axis, so that it
would have a supersonic motion at peri- or apoastron only for $e >
2h$. This implies that the upper part of the $\dot e$ and $\dot a$
maps presented in section~\ref{sec:fiducial-calculation} may be
different, but not the main part, for $e<0.1$ (our disc has
$h=0.05$). In particular, all the fixed points we have found have an
eccentricity below $2h$ (see Figs.~\ref{fig:fixedr}
and~\ref{fig:fixedl}), so they should not be affected by this effect.

One dimensional calculations of the structure of a pebble accreting
planetary envelope show that pebbles do not hit directly the core
above a fraction of an Earth mass
\citep{2018A&A...611A..65B}. Instead, they vaporise before reaching
the core and form a high metallicity envelope around the core. The
energy release is then smaller than that given by
Eq.~\eqref{eq:42}. The reduction factor for the luminosity that we
introduced in section~\ref{sec:vari-from-fiduc} is an \textit{ad hoc}
attempt to take this effect into consideration. However, further
complications come into play: part of the envelope may be recycled by
gas flowing in and out of the Bondi sphere
\citep{2015MNRAS.447.3512O,2015arXiv150503152F}, which may limit the
core's growth \citep{2021A&A...653A.103B}. Besides, the rate of
recycling increases with the planetary eccentricity
\citep{Bailey_2021}. Much further work is therefore needed to assess
to which extent an eccentric planet retains it high metallicity
envelope.  Regardless of the planet's eccentricity, results obtained
from calculations with a steady flow of pebbles may differ
substantially from results obtained when the flow of pebbles is
intermittent, as is the case for the mechanism presented here. In
addition, our numerical scheme implies that the heat released by
accretion is injected instantaneously in the eight cells surrounding
the planet. While the diffusion timescale of the heat within the hot
trail is a small fraction of the orbital timescale in the headwind
regime \citep{2017arXiv170401931E} and is properly accounted for by
our numerical scheme, the delay of heat transfer from the planet to
the immediately surrounding cells involves several phases: the
emergence of heat from the convective zone, and its subsequent
transfer by radiation up to the distance of the neighbouring cell
centres. Assessing this delay should be done with a specific,
time-dependent study of pebble accretion down to the core scale.

  Finally, our study cannot capture properly the dynamics of the
  planet when its mass exceeds the pebble isolation mass (PIM). The
  latter indeed requires that a local pressure maximum be created on
  the outer side of the orbit. However, the time it takes for this
  maximum to appear is larger than the 3-orbit duration of our
  runs. We comment that the PIM has been studied for the case of
  eccentric planets \citep{2022MNRAS.510.3867C} when there is a steady
  flow of pebbles originating from the outer disc. Here, there is
  rather a given reservoir of pebbles already in the ring from which
  the planet accretes. Whether and how accretion proceeds once the
  planet's mass exceeds the PIM should be the subject of future
  studies. Also, we mention that \citet{2021MNRAS.503L..67S} found
  that the PIM is significantly increased at pressure bumps, so the
  largest mass of our numerical study may well be significantly below
  the isolation mass.

\subsubsection{Other limitations}
\label{sec:other-limitations}
In addition to the caveats listed above, we mention that our
analysis considers only one value for the Stokes number
($\tau_s=0.01$) and one value for the turbulence parameter
($\alpha_\nu=10^{-4}$). \citet{2024arXiv240205760P} consider rings with
  different Stokes number and find that the accretion spikes near
  perihelion are shorter and higher when $\tau_s=0.1$, as can be
  expected for the more narrow dusty rings obtained in that case. They
  find that planets in that case as subjected to a similar trapping
  than those  of the case $\tau_s=0.01$, but have a mass that saturate
  at smaller values. They also perform simulations with
  $\tau_s=10^{-3}$, in which the ring has a width marginally larger than
  the aspect ratio of  the gas, which show a trapping similar to that obtained with
  larger Stokes numbers, but for which planets undergo a much slower
  growth.

We also add neither the mass nor the momentum accreted from the dust
to the planet. Over the very short timescales of our run, the mass
accreted is extremely small and neglecting it is legitimate. As for
the momentum of the dust, we expect that it would change in a sizeable
manner the orbital parameters of the planet when the mass accreted is
comparable to the mass of the planet. Since the timescales for the
evolution of eccentricity are at least one order of magnitude shorted
than the mass doubling times, the impact of the accreted momentum on
the evolution of eccentricity should be subdominant compared to that
of the thermal forces.

\section{Conclusions}
\label{sec:conclusions}
We study the orbital evolution of a planetary embryo forming in a
dusty ring of $10\;M_\oplus$ at $10$~au from a solar mass star. We
take into account the accretion of dust and the radiative feedback on
the gaseous disc: the energy released by accretion is used to heat the
nearby gas. Even at the starting mass of our study ($0.1\;M_\oplus$),
embryos are found to be sufficiently luminous to have their
eccentricities excited. After undergoing eccentricity growth, they end
up trapped in one of these two stable orbits: an orbit outside the
ring, with a periastron in the ring, at which the planet accretes at
each passage, or an orbit inside the ring, with its apoastron slightly
interior to the ring, so that the planet accretes significantly less
on this orbit than on the first one. A critical mass core can be
assembled in a fraction of a Myr on the outer orbit, while the core
may remain subcritical over the disc lifespan on the inner orbit. The
eccentricity of the outer orbit is comparable to the aspect ratio of
the gaseous disc and the eccentricity of the inner orbit is smaller,
typically by a factor of two. Which of the two orbits is eventually
reached by growing cores depends on the starting point of the embryo
in the ring. The watershed between inner and outer tracks is close to
the peak of dust density, hence an accurate value of the torque
exerted on a low-mass embryo prior to its eccentricity growth is
required to determine whether it will eventually join the inner or the
outer orbit. Our model tends to favour the outer orbit, but it is
likely that the preferred orbit depends on the specifics of the ring,
or even on the exact time at which the growth of eccentricity starts,
as the torque exerted by the dust and gas in a low-viscosity disc may
have a stochastic component.  The trapping at the outer point relies
on a vigorous heating force at periastron, arising from the high
accretion rate of the planet on this portion of its orbit. This force
has same direction as the planet's motion, and increases the planet's
angular momentum and orbital energy. This increase at periastron
compensates the variation of these two quantities over the rest of the
orbit. The variation of the planet's luminosity as a function of the
orbital phase is therefore a key ingredient for the trapping on the
outer orbit. The trapping at the inner point is also based on thermal
forces, but details differ: the planet adopts the eccentricity at
which its migration changes sign. At an adequate distance from the
ring, this eccentricity turns out to be constant in time. A much
milder modulation of the planet's luminosity as a function of the
orbital phase is required for a trapping on the inner orbit, which is
why the planet never approaches the dust as much as on the outer
orbit.  There is a considerable leeway for our mechanism to
operate. Even when we arbitrarily reduce the accretion luminosity by a
factor of ten, we still find an outer and inner eccentric trapping
orbit. Since the existence and location of the outer orbit is
precisely based on the accretion of dust, the planet is \textit{de
  facto} a dust hunter, and consumes the ring in a systematic fashion.
The trapping mechanism at the outer orbit does not rely on the existence
of a pressure bump (we did set up one, to give rise to a dust ring,
but this is not necessary for the trap to exist). It occurs whenever the dust has a
radially narrow distribution. It should therefore occur in the clumpy
rings described by \citet{2021MNRAS.505.1162J}. An interesting
prospect is that these rings can expand with time. As they do, so does
the outer trapping orbit: these rings can bring forming planets to
several tens of astronomical units over Myr timescales.  The present
work suggests that as long as protoplanets in the Earth-Neptune mass
range are accreting, even weakly so, their eccentricities are driven
by the disc rather than damped. \citet{2022MNRAS.517.4472L} find
evidence for such driving in the apsidal alignment of resonant
transiting pairs.  A necessary and significant step forward to make
more accurate predictions about the accretion of dust by an eccentric
planet would be a small scale description of pebble accretion, down to
the core, in a time varying flow with a time varying pebble
input. This would allow us to better estimate the value of the planet's
luminosity.

\section*{Acknowledgements}
The authors wish to thank the referee, R. O. Chametla, for comments
that led to an improvement of this manuscript, and O. Chrenko for his
insightful feedback. The simulations included in this work were executed
on the Stellar and Della clusters at Princeton University as well as
the Piz-Daint cluster at CSCS under the project
s1077. F.~M. acknowledges support from UNAM's grant PAPIIT~107723,
UNAM's DGAPA PASPA program and the Laboratoire Lagrange at
Observatoire de la Côte d'Azur for hospitality during a one-year
sabbatical stay.  P.~B.~L. acknowledges support from ANID, QUIMAL fund
ASTRO21-0039 and FONDECYT project 1231205.

\section*{Data Availability}
The FARGO3D setup used in the present work will be shared upon reasonable request to the corresponding author.



\bibliographystyle{mnras}
\bibliography{biblio}



 \appendix
 \section{Average thermal torque on a planet of fixed luminosity}
 \label{sec:aver-therm-torq}
We assume that the corotation offset is a small fraction 
$\epsilon$ of the epicyclic excursion $ea$: $x_p^0=\epsilon ea$, where
$|\epsilon| \ll 1$ and rewrite the coordinates of the velocity
w.r.t. the gas of Eq.~\eqref{eq:14} as: 
\begin{equation}
  \label{eq:54}
  (\dot x|_{\rm gas},\dot y|_{\rm gas}) =
  \Omega_pea\left[\sin(\Omega_pt),\frac 12\cos(\Omega_p t)+\frac 32\epsilon\right]
\end{equation}
The unit vector $\mathbf{n}$ with same direction as the planet's velocity with respect to the 
gas has therefore the components: 
\begin{equation}
  \label{eq:55}
  (n_x,n_y) =A\times \left[\sin(\Omega_pt),\frac 
     12\cos(\Omega_pt)+\frac 32\epsilon\right].
\end{equation}
with:
\begin{equation}
  \label{eq:56}
  A = \frac{1}{\sqrt{\sin^2(\Omega_p t)+\left[\frac 
        12\cos(\Omega_pt)+\frac32\epsilon\right]^2}}
\end{equation}
Using the fact that $|\epsilon| \ll 1$, we expand the prefactor of Eq.~\eqref{eq:55}
as: 
\begin{equation}
  \label{eq:57}
  A=\frac{1}{\left[1-\frac 34\cos^2(\Omega_pt)\right]^{1/2}}-
  \frac{\frac 34\epsilon\cos(\Omega_pt)}{\left[1-\frac34\cos^2(\Omega_pt)\right]^{3/2}}
\end{equation}
The thermal force exerted on the planet has expression: 
\begin{equation}
  \label{eq:58}
  \mathbf{F} = F_d\mathbf{n}, 
\end{equation}
and (the vertical component of) its torque is: 
\begin{equation}
  \label{eq:59}
  \Gamma=(a+x)F_y-yF_x.
\end{equation}
Using Eqs.~\eqref{eq:4}, \eqref{eq:55}, \eqref{eq:57} and~\eqref{eq:59}
we obtain:
\begin{equation}
  \label{eq:60}
  \begin{split}
  \Gamma&=F_da\left\{[1-e\cos(\Omega_pt)]\left[\frac 
      12\cos(\Omega_pt)+\frac 32\epsilon\right]\right\}\\
  &\left.\phantom{\frac 32}-2e\sin^2(\Omega_pt)\right\}\\
  &\times \left\{
    \frac{1}{\left[1-\frac 34\cos^2(\Omega_pt)\right]^{1/2}}-
  \frac{\frac
    34\epsilon\cos(\Omega_pt)}{\left[1-\frac34\cos^2(\Omega_pt)\right]^{3/2}}
  \right\}
  \end{split}
\end{equation}
We average this expression over one orbital period. Retaining only
terms to first order in $e$ and $\epsilon$ that do not cancel out, we obtain: 
\begin{equation}
  \label{eq:61}
  \begin{split}
  \langle\Gamma\rangle=&F_dae\left\langle\frac{-\frac 
      12\cos^2(\Omega_pt)-2\sin^2(\Omega_pt)}{\left[1-\frac 
        34\cos^2(\Omega_pt)\right]^{1/2}}\right\rangle \\
  &+F_da\frac 32\epsilon\left\langle \left[1-\frac 34\cos^2(\Omega_pt)\right]^{-1/2}\right\rangle\\
 & -F_da\frac 38\epsilon \left\langle  \frac{\cos^2(\Omega_pt)}{\left[1-\frac 
        34\cos^2(\Omega_pt)\right]^{3/2}}\right\rangle. 
  \end{split}
\end{equation}
Each of the averages in the above expression can be expressed in terms
of the complete elliptic integrals of the first and second kind
(denoted respectively ${\cal K}$ and ${\cal E}$).  The average factor
of the first line is $-4{\cal E}(-3)/(2\pi)\approx -1.5420$. That of
the second line is $8{\cal K}(-3)/(2\pi)\approx 1.3729$ and that of
the third line is
$(32/3)[{\cal E}(-3)-{\cal K}(-3)]/(2\pi)\approx 2.2814$.  We
therefore have the following expansion:
\begin{equation}
  \label{eq:62}
  \langle\Gamma\rangle=F_da(1.2038\epsilon -1.5420e), 
\end{equation}
which can be recast as Eq.~\eqref{eq:16}.

\section{Average thermal torque on a planet with variable luminosity}
\label{sec:aver-therm-torq-1}
Using the dependence of Eq.~\eqref{eq:18}, we have an extra factor
$[1-es\cos(\Omega_pt)]$ for the $x-$ and $y-$components of the thermal
force, hence for the torque, which has therefore the expression:
\begin{equation}
  \label{eq:63}
  \begin{split}
      \Gamma&=F_da \left[1-es\cos(\Omega_pt)\right]\left\{[1-e\cos(\Omega_pt)]\left[\frac 
    12\cos(\Omega_pt)+\frac 32\epsilon\right]\right.\\
&\left.\phantom{\frac 32}-2e\sin^2(\Omega_pt) \right\}\\
&\times \left\{\left[1-\frac 34\cos^2(\Omega_pt)\right]^{-1/2}-\frac 34\epsilon\frac{\cos(\Omega_pt)}{\left[1-\frac34\cos^2(\Omega_pt)\right]^{3/2}}\right\}
  \end{split}
\end{equation}
The time average of this quantity to first order in $e$ and $\epsilon$, in addition to the terms obtained
in Eq.~\eqref{eq:61}, contains the following term, which scales with $s$:
\begin{equation}
  \label{eq:64}
  \langle\Gamma\rangle_s=-F_da\frac{es}{2}
  \left\langle\frac{\cos^2(\Omega_pt)}{\left[1-\frac
        34\cos^2(\Omega_pt)\right]^{1/2}}\right\rangle
\end{equation}
The average factor in this equation can be cast as $(8/3)[4{\cal
  K}(-3)-{\cal E}(-3)]\approx 0.8025$, hence
 \begin{equation}
   \label{eq:65}
   \langle\Gamma\rangle_s=-0.40F_daes,
 \end{equation}
from which we infer Eq.~\eqref{eq:19}.

\bsp	
\label{lastpage}
\end{document}